\documentclass[11pt]{article}

\usepackage[utf8]{inputenc}
\usepackage[T1]{fontenc}
\usepackage[a4paper,margin=1in]{geometry}
\usepackage{graphicx}
\usepackage{xcolor}
\usepackage{hyperref}
\usepackage{orcidlink}
\usepackage{microtype}
\microtypesetup{expansion=false}
\usepackage{amsmath}
\usepackage{amssymb}
\usepackage{latexsym}
\usepackage{float}
\usepackage{multirow}
\usepackage{makecell}
\usepackage[mathscr]{euscript}
\usepackage{mathrsfs}
\usepackage{subfigure}
\usepackage{stackrel}
\usepackage{tabularx,booktabs}
\usepackage{pifont}
\usepackage{array}
\usepackage{adjustbox}
\usepackage{ragged2e}
\usepackage{xspace}
\usepackage{slashed}
\usepackage{lineno}

\graphicspath{{Images/}{./Images/}{./}}

\hypersetup{citecolor=blue,linkcolor=blue,urlcolor=blue,colorlinks=true}
\colorlet{blue}{black}

\newcolumntype{Y}{>{\centering\arraybackslash}X}
\newcolumntype{R}[1]{>{\RaggedRight}p{#1}}

\def\be{\begin{equation}}
	\def\ee{\end{equation}}
\def\bi{\begin{itemize}}
	\def\ei{\end{itemize}}
\def\ben{\begin{enumerate}}
	\def\een{\end{enumerate}}

\def\apj{Astrophys. J.}
\def\apjl{Astrophys. J. Lett.}
\def\apjs{Astrophys. J. Supp. Ser. }

\def\physrep{Phys. Rep. }
\def\mnras{Mon. Not. Roy. Astron. Soc. }

\def\prd{Phys. Rev. D.}

\def\pasa{Proc. Astron. Soc. Pac.}

\begin{document}
	
	\title{Classifying the nuclear equation of state in LVK interferometric noise through core-collapse supernova gravitational-wave signatures using convolutional neural networks}
	
	\author{%
		Alejandro~Casallas-Lagos$^{1}$\orcidlink{0000-0001-5985-8819}\thanks{Corresponding author: \href{mailto:a.casallas-lagos@uw.edu.pl}{a.casallas-lagos@uw.edu.pl}},
		Marek J. Szczepa\'nczyk$^{1}$\orcidlink{0000-0002-6167-6149},\\
		Michele~Zanolin$^{2}$,
		Anthony~Mezzacappa$^{3}$,
		Javier~M.~Antelis$^{4}$,\\
		Daniel~Murphy$^{3}$,
		and Claudia~Moreno$^{5}$\\[0.75em]
		\footnotesize $^{1}$University of Warsaw, Faculty of Physics, Institute of Theoretical Physics, Warsaw, Poland\\
		\footnotesize $^{2}$Department of Physics and Astronomy, Embry-Riddle Aeronautical University, Prescott, AZ 86301, USA\\
		\footnotesize $^{3}$Department of Physics and Astronomy, University of Tennessee, Knoxville, TN 37996-1200, USA\\
		\footnotesize $^{4}$Tecnol\'ogico de Monterrey, Escuela de Ingenier\'ia y Ciencias, Monterrey, N.L., 64849, M\'exico\\
		\footnotesize $^{5}$Departamento de F\'isica, Universidad de Guadalajara, Guadalajara, Jal., 44430, M\'exico
	}
	\date{}
	
	\maketitle
	
	\begin{abstract}
		This paper presents a convolutional neural network (CNN) approach to classifying the nuclear equation of state (EOS). We use as illustrative examples five two-dimensional simulations of core collapse supernovae (CCSN) varying only the EOS. We use the estimation of the high-frequency feature (HFF) initial slope in real interferometric data from the O3b LVK scientific run at Galactic distances of 1 kpc, 5 kpc, and 10 kpc. 
			Our CNN model classifier achieves an overall accuracy of 98.58\% at a distance of 1 kpc and 52.43\% at 5 kpc for each class. This accuracy progressively declines at an extended distance of 10 kpc, where the capacity of the classifier vanishes.
			The model’s success in EOS differentiation at 1 kpc suggests scalability for next-generation observatories; we expect the 
			order-of-magnitude sensitivity improvements of Cosmic Explorer and the Einstein Telescope to enable similar classification performance at ten times the current distance.
			The implementation of more specific performance metrics, such as the macro-average one-vs-rest (OvR) area under the curve (AUC), reveals an overall macro-average OvR AUC of 0.97 and 0.98 at 1 kpc. This performance stands both collectively across all classes and individually.
	\end{abstract}
	
	\noindent\textbf{Keywords:} core-collapse supernovae; gravitational waves; nuclear equation of state; convolutional neural networks; Coherent WaveBurst; LVK detector noise
	
	\section{Introduction}
	\label{sec:Introduction}
	The first detection of a gravitational wave (GW) by LIGO~\cite{Aasi_ALIGO_2015} in September 2015, event GW~150914 \cite{Abbott_GWBH_2016}, gave birth to GW astronomy. Since then, the collaborative efforts of LIGO, VIRGO~\cite{Acernese_2014}, and KAGRA~\cite{Aso_KAGRA_2013} LVK collaboration have validated the detection of more than 400 events related to compact binary systems \cite{Abbott_GWTC2_2023, Abbott_GWTC3_2023, LVK_GWTC-4.0_2025}. This remarkable contribution has motivated the improvement of detector network sensitivity to explore new, non detected, sources of GWs, such as those produced by core-collapse supernovae (CCSNe) \cite{Powell:2023bex}. 
	CCSNe are multimessenger astronomy (MMA) systems characterized by the emission of different carriers of physical information; electromagnetic radiation (photons), neutrinos, and GWs, making them exceptional laboratories that contain diverse sources of physical information, all coexisting in a single event \cite{Kuroda_2018, Radice_2019, JaMeSu16, MezzZan_2024, M_ller_16}.
	\\ \\
	Despite the low estimated rate of core-collapse supernova (CCSN) events in our Galaxy—approximately two per century \cite{Burrows_2021, Abdikamalov_SNHB_2022}—significant progress has been made in preparing for the inaugural detection of a Galactic or near-Galactic CCSN GW event. This extensive preparatory work spans diverse fields, encompassing theoretical investigations into the nuclear equation of state (EOS) for dense matter and various explosion mechanisms, including the neutrino-driven mechanism \cite{Melson_2015, Mezz_Endv_2020, Janka_2012, Mezzacappa_2023, Murphy_2009, Takiwaki_2018}. 
	At the same time, dedicated software has been developed to detect and characterize burst transients and to implement noise reduction techniques in LVK interferometric data \cite{Klimenko_2005, Klimenko_2008, Klimenko_2016, Klimenko_2022, Szczepa_czyk_2021, Mukherjee_2017, Mukherjee_2021, Antelis_2022, Summerscales:2007xq, Logue:2012zw, LopezPortilla:2020odz}, alongside advances in observational and theoretical techniques in asteroseismology analysis \cite{Aerts_AST_2021, Handler_2013, Kurtz_AST_2022, Rodriguez_AST_2023, Torres-Forne_AST_2017, Torres-Forne_AST_2019, Murphy_2025, Westernacher_2019, Westernacher_2020}. 
	Furthermore, robust CCSN numerical simulations have played a crucial role in modeling these complex phenomena \cite{Kuroda_2016, Kuroda_2018, Mezzacappa_2020, Mezz_D_GW_2023, Mezz_Endv_2020, Andresen_2019, Andresen_2017, M_ller_2012, M_ller_2013, M_ller_2017, M_ller_2020, Hanke_2013, Melson_2015, Vartanyan_2018, O_Connor_2018, Powell_2019, Powell_2020, Yoshida_2021, Murphy_2009, KuFiTa22, KoIwOh09, KoIwOh11, KoSaKa06, VaBuRa19, WoHe07, CoWhi_1966, JaMeSu16}, complemented by significant breakthroughs in laser interferometry that enhance GW detector capabilities \cite{Abbott_O3_2023}. 
	The future detection of GW from CCSNe promises to deliver a wealth of information essential for understanding their explosion mechanisms. Achieving such a detection would stand as a monumental success for the LVK collaboration, marking a significant leap forward in the field of GW physics and MMA.
	\\ \\   
	\textcolor{blue}{Numerical simulations of CCSNe from GW signals are inherently stochastic due to the turbulent and nonlinear dynamics of the post-bounce phase. Nevertheless, their time–frequency evolution exhibits robust and reproducible features that can be treated as quasi-deterministic. One of the most prominent of these features is what we referred to as the high frequency feature (HFF) (see \cite{MezzZan_2024} for a discussion of the terminology). In time–frequency spectrograms, this feature appears as a continuous, upward-trending track, typically emerging at frequencies of $\sim 250$–$400$ Hz at $\sim 100$ ms after core bounce and increasing up to $\sim 1$–$2$ kHz as the PNS evolves.
		Recent studies have shown that the HFF arises from a combination of PNS oscillation modes, including both $f$- and $g$-modes, whose relative contributions evolve over time.} 
	A significant challenge in GW astronomy and MMA is accurately estimating the astrophysical parameters of GW sources using ground-based GW detectors data \cite{MezzZan_2024}. Recent research shows that the high-frequency feature (HFF) can be correlated with progenitor's physical properties of the emitted GWs \cite{M_ller_2012, M_ller_2013, Morozova_2018, Couch_2013, Torres-Forne_AST_2017, Torres-Forne_AST_2019, Murphy_2025,2bdz-783t} providing a framework for using interferometric data to understand the physical characteristics of the GW sources. 
	\\ \\
	Our prior investigations \cite{Casallas_2023, Murphy_Casallas_2024} laid the groundwork for the current study, focusing on two key areas:
	(i) HFF Initial Slope Estimation in LVK noise: we developed a machine learning (ML) methodology to estimate the dominant linear component of the High-Frequency Feature (HFF) slope embedded in CCSN GW signals. This method uses GW events reconstructed with Coherent WaveBurst (cWB) \cite{Klimenko_2005, Klimenko_2016, Klimenko_2008, Drago_2021}, a software designed for detecting and reconstructing GW bursts with minimal assumptions about signal morphology. 
	Our analysis characterized the feasibility of training a ML algorithm \cite{Chan:2019fuz, Abylkairov:2024hjf, Mitra:2022nyc, Iess:2023quq} to estimate the HFF initial slope in real interferometric LVK noise  and across different Galactic distances. We found that the estimation of the HFF slope is influenced by several Proto-Neutron Star (PNS) parameters, including mass, angular velocity, EOS, and progenitor metallicity. 
	(ii) The impact of the EOS in the HFF slope estimation: We conducted a detailed examination of how variations in the EOS influence HFF slope estimation within the noise environment of the LVK interferometers, investigating EOS variations in the absence of rotation and magnetic fields (see \cite{Murphy_Casallas_2024}, Section V, Tables II and III for details).
	\\ \\
	Building on our previous research \cite{Casallas_2023, Murphy_Casallas_2024} this study presents a systematic investigation of the influence of EOS on HFF initial slope estimation by isolating EOS variations while maintaining constant progenitor parameters.
	We use a methodology based on an image recognition model, specifically a Convolutional Neural Network (CNN), designed to distinguish between five EOSs, specifically DD2, FSUgold, IUSFU, SFHo, and SFHx, for the \textit{Chimera} E-series at three Galactic distances of 1 kpc, 5 kpc and 10 kpc in real interferometric noise of the LVK scientific run O3b \cite{2023ApJS..267...29A}.
	Our methodology involves a three-phase approach to evaluate the EOS classification performance. First, we trained and tested our CNN model for EOS multi-class classification using data derived from cWB event production. This initial dataset is extracted from a one-week Time Window (TW1) during the second half of the LVK scientific run O3b. Second, we independently trained and tested the CNN model using data from a second, one-week Time Window (TW2), acquired one week after TW1. Finally, to assess the CNN model's robustness and efficiency against temporal variations within LVK noise, we trained the CNN on data from TW1 and evaluated its performance using data from TW2.
	In future works we plan to include also the role of progenitor mass \cite{Mezz_D_GW_2023, Vartanyan_2018}, rotation rate \cite{Andresen:2018aom, Pajkos_2019, Pan_2018, Yoshida_2021, Pastor-Marcos:2023tcc}, and magnetic fields \cite{Pan_2021, Obergaulinger_2021}.
	\\ \\
	\textcolor{blue}{Our method analyzes input images from cWB events using a CNN multi-classifier algorithm \cite{CNN1, CNN2}. Inspired by the human visual cortex, these architectures are highly effective for image classification and object detection tasks. While Receiver Operating Characteristic (ROC) curves are traditionally employed for signal detection purposes, they have recently been implemented for the astrophysical interpretation of CCSN data \cite{Lin_2023}. Here, we introduce them as a metric for evaluating EOS classification performance, where each operating point on the curve is obtained by varying the discrimination threshold of the CNN's output probabilities. This framework, alongside the methodology for HFF initial slope estimation presented in \cite{Casallas_2023, Murphy_Casallas_2024}, enriches the suite of computational tools for contemporary CCSN gravitational-wave parameter estimation, offering expanded avenues to investigate the fundamental physics encoded within these astrophysical signals.
		The goal of the CNN algorithm implementation is to assign class labels correlated with a specific EOS HFF slope value for each Galactic distance considered in this study.} 
	\\ \\
	The manuscript is organized as follows. 
	Section \ref{sec:EOS-HFF} presents numerical evidence for the correlation between the estimation of the HFF slope and the EOS in LVK data. Section \ref{sec1:physics} provides an overview of recent advances detailing the link between the estimation of the HFF slope and the progenitor parameters. Section \ref{sec:Methodology} outlines the different stages of constructing the CNN algorithm aimed at EOS classification, which is based on the estimated HFF derived from cWB-XP event production, as well as the architecture, evaluation, and validation of the classifier. In section~\ref{sec:results}, we deliver the findings from the implementation and validation of the CNN model. The investigation into the CNN architecture's validation focuses on evaluating the algorithm's performance in classifying the EOS for three Galactic distances. To evaluate the accuracy of the CNN classifier, we provide a set of performance metrics, including the receiver operating characteristic for the One-vs-Rest multiclass scheme and the associated confusion matrices. Finally, section~\ref{sec:summary} presents the conclusions and future research directions for this work.
	%
	\section{HFF slope estimation and the EOS}\label{sec:EOS-HFF}
\begin{figure*}[!ht]
		\centering
		\includegraphics[width=0.3\textwidth]{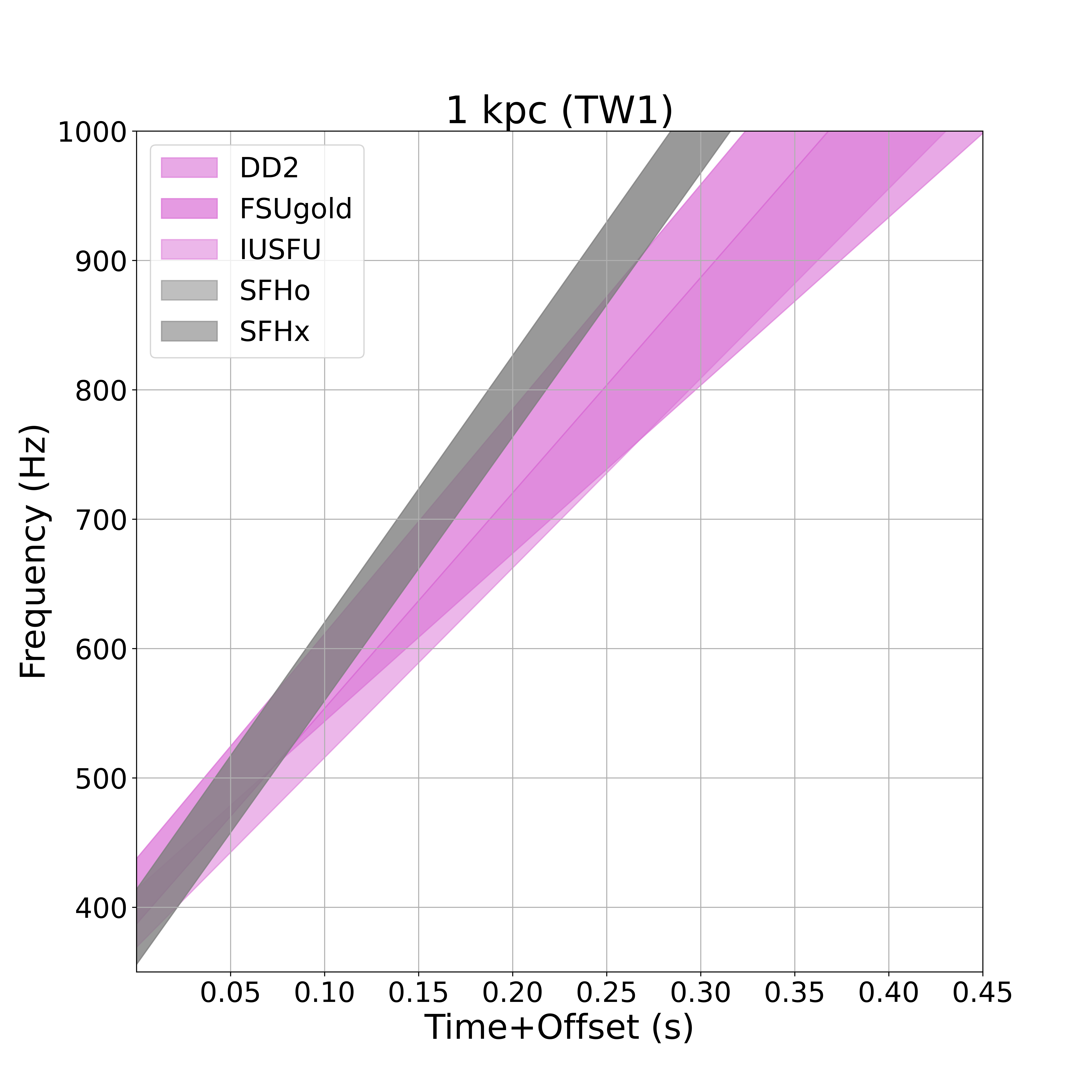}
		\includegraphics[width=0.3\textwidth]{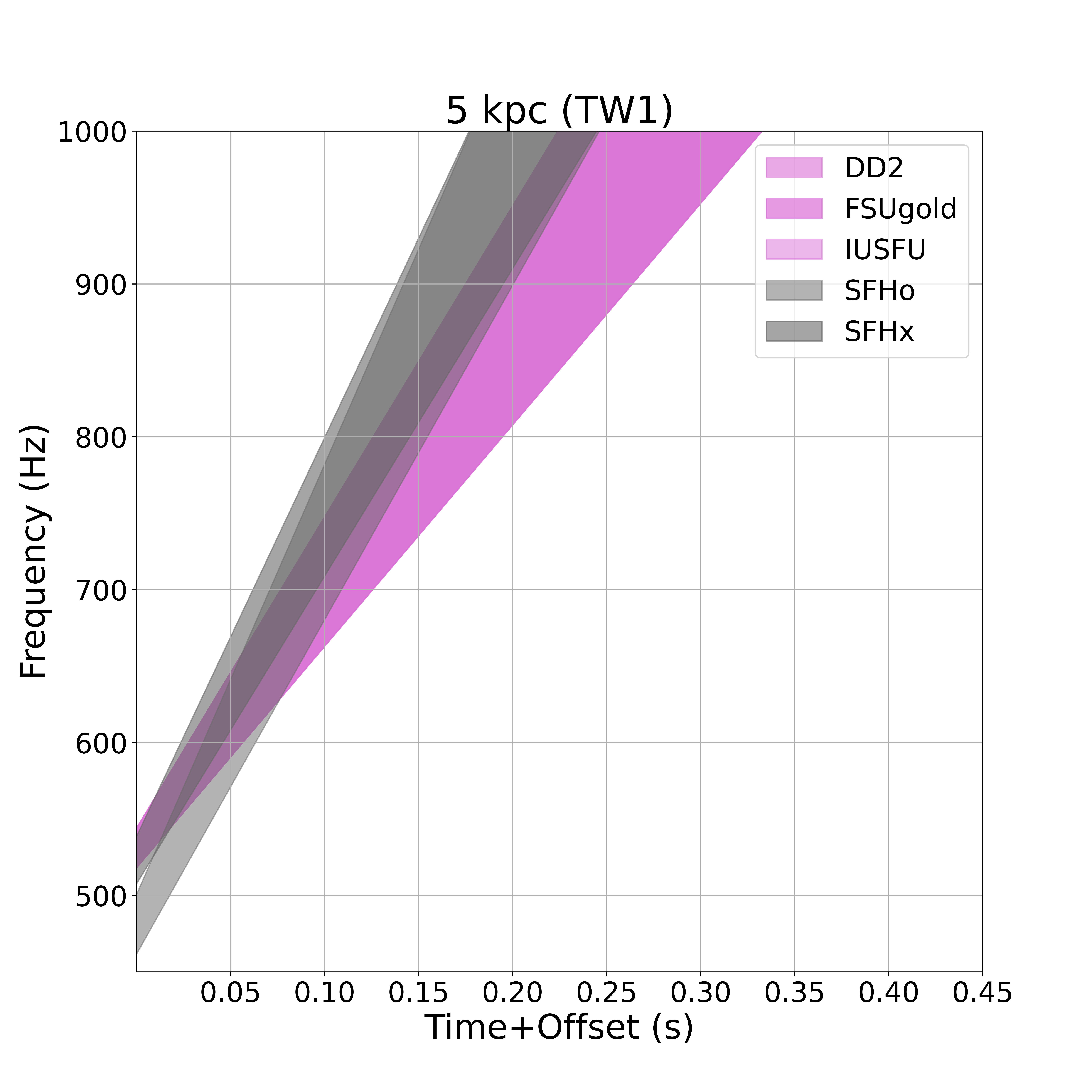}
		\includegraphics[width=0.3\textwidth]{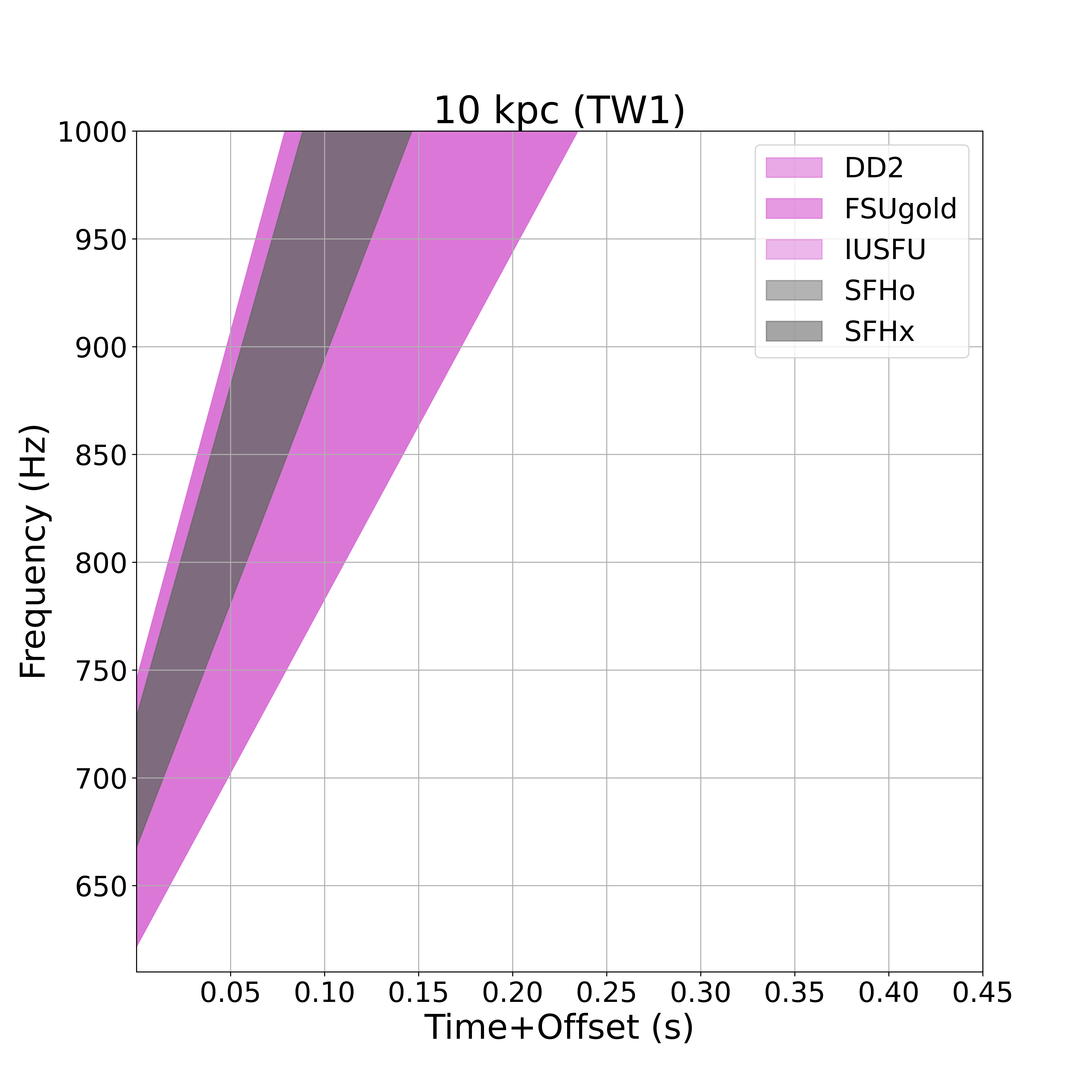}
		\includegraphics[width=0.3\textwidth]{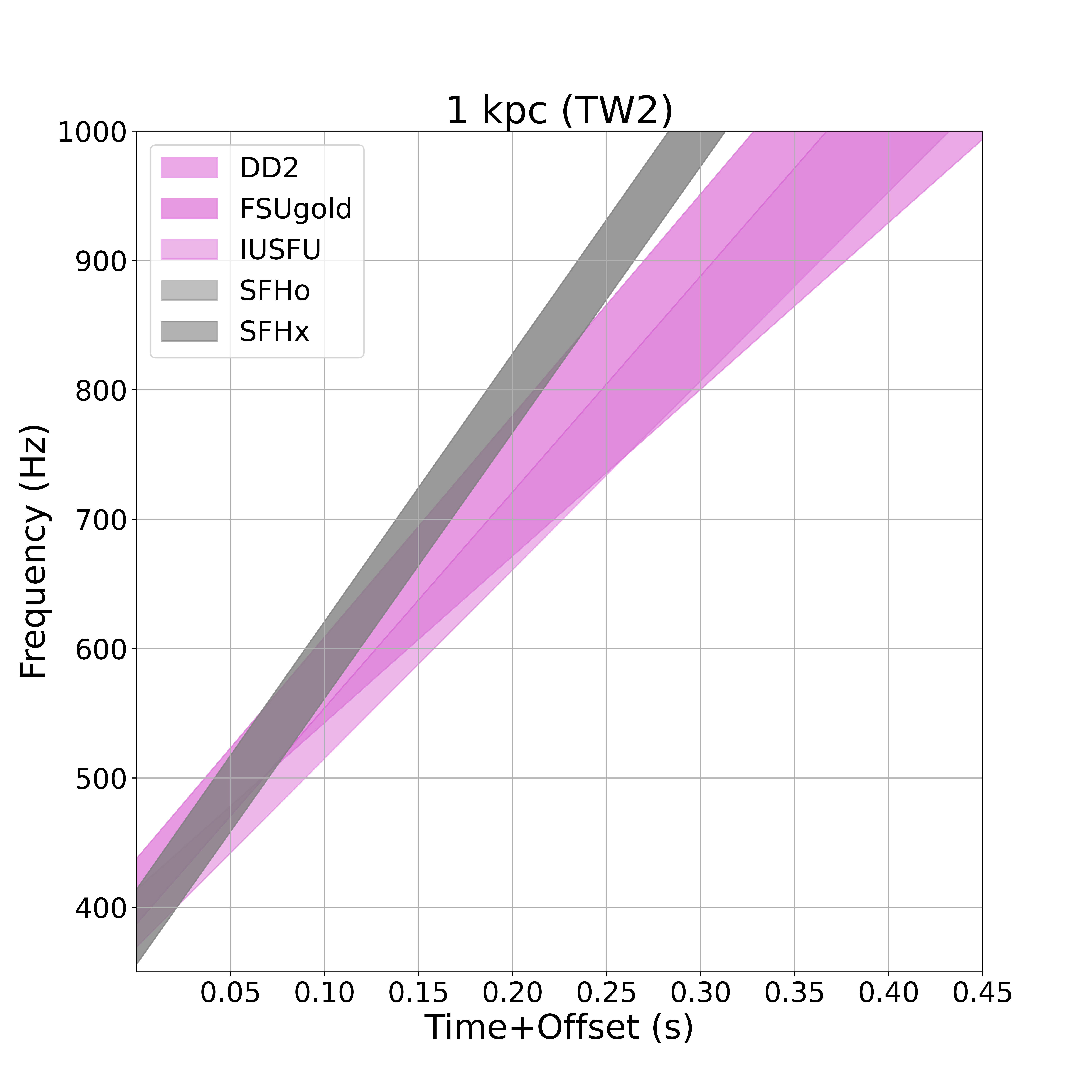}
		\includegraphics[width=0.3\textwidth]{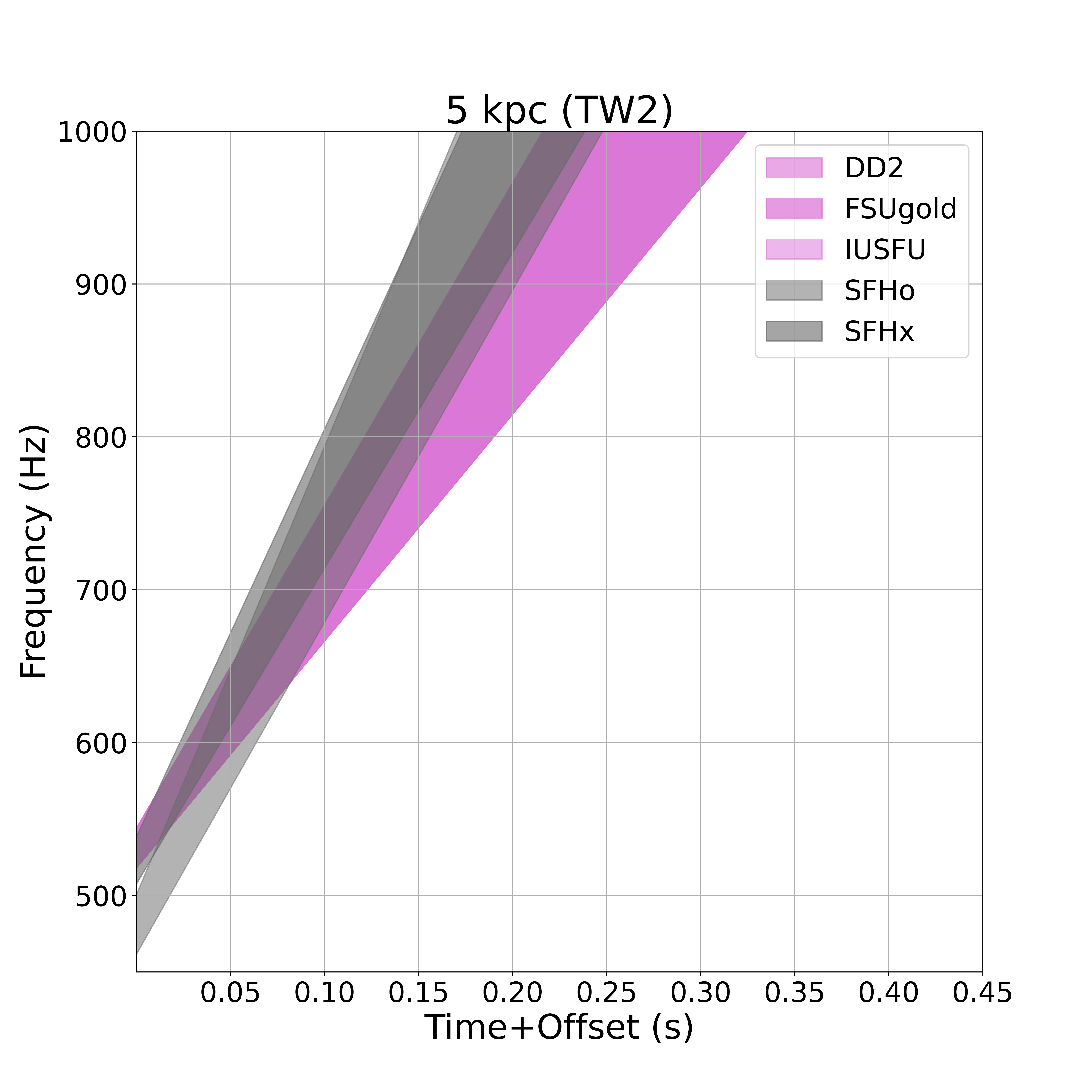}
		\includegraphics[width=0.3\textwidth]{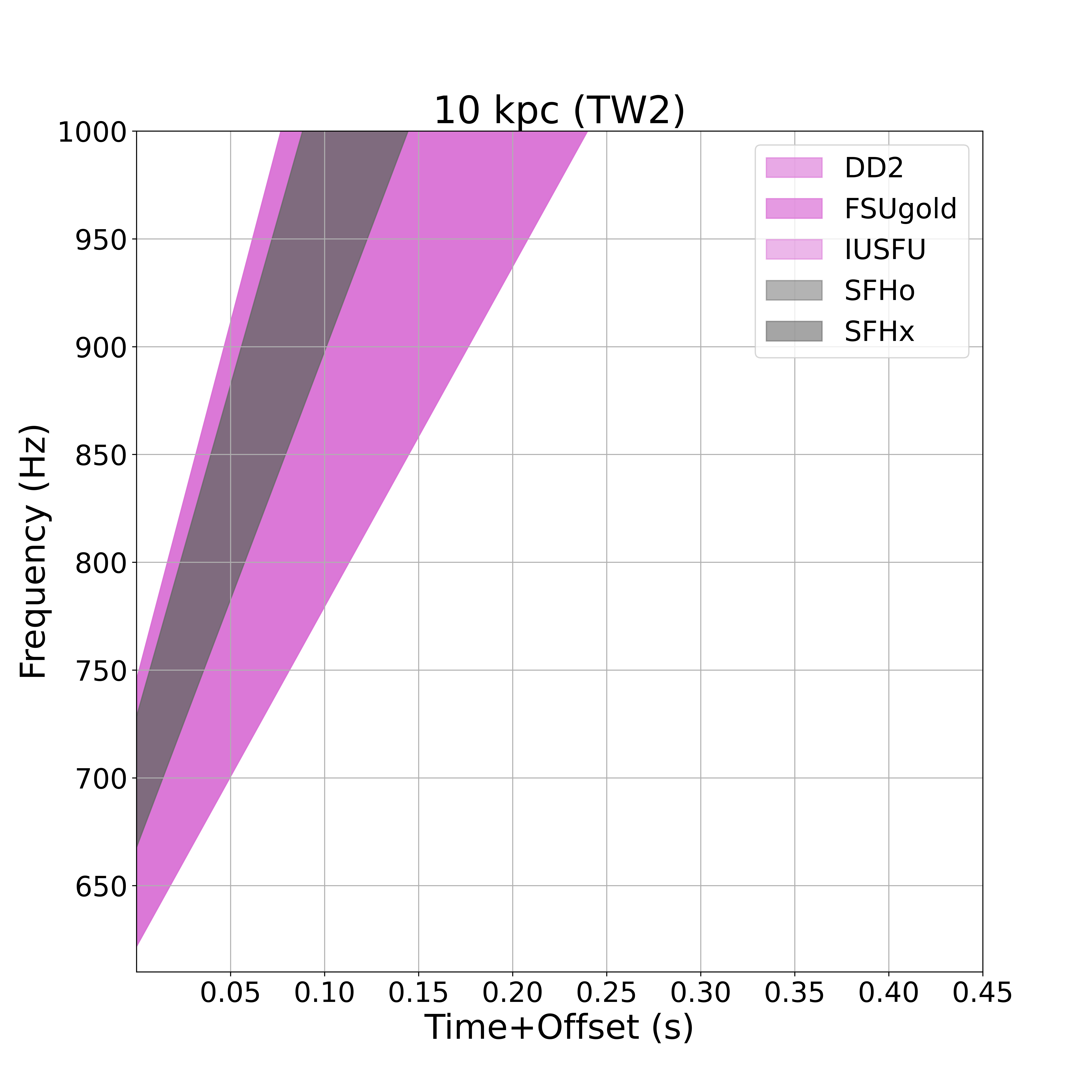}
		\caption{Range of variability of the estimated HFF slopes for the \textit{Chimera} E-series CCSN GW signals. These signals, described in \cite{Landfield_thesis}, are used to estimate the HFF slope in real LVK data across five different EOS: DD2, FSUgold, IUSFU, SFHo, and SFHx, with the full analysis detailed in \cite{Murphy_Casallas_2024}.
			The figure shows that at a Galactic distance of 1 kpc, the predicted HFF slopes, associated with each EOS, naturally arrange into two primary clusters: a gray group comprising SFHo and SFHx, and a purple/violet group consisting of DD2, FSUgold, and IUSFU. 
			The analysis is performed using the independent TW1 and TW2 datasets described in Table \ref{tab:Time_Window}. The top panel shows the results associated with both TW1 and TW2 across 1 kpc, 5 kpc, and 10 kpc. 
			At 1 kpc the two EOS groups are clearly separable. However, as the Galactic distance increases, this clear distinction progressively fades, leading to a merger of the clusters. This merging trend suggests a growing difficulty in differentiating between individual EOS based on their estimated HFF slopes at greater distances. 
			The figure visually confirms the convergence of the two main EOS groups until they become indistinguishable at 10 kpc, reinforcing the conclusions drawn in \cite{Murphy_Casallas_2024}.}
		\label{fig:scatter}
	\end{figure*}

	%
	Tables \ref{tab:HFF_TW1} and \ref{tab:HFF_TW2} illustrate the noise-free estimated HFF initial slope ($s$) and the mean estimated slope ($\hat{\bar{s}}$) derived from LVK strain data from the second part of the third observing run (O3b). Additionally, for each detection distance, we report the standard deviation (STD) and root-mean-square error (RMSE) associated with ($\hat{\bar{s}}$). These latter metrics serve as a measure of agreement between the noise-free HFF slope and the mean slope estimated by the DNN across different noise realizations. 
	\\ \\
	The study presented in this paper is based on GW signals generated from two-dimensional, axisymmetric CCSN simulations conducted with the \textit{Chimera} code \cite{Bruenn_2020}. Our analysis incorporates five specific EOS models (DD2, FSUGold, IUFSU, SFHo, SFHx) that dictated the nuclear physics in these simulations, and will be collectively designated as the CCSN GW \textit{Chimera} E-series throughout this paper. The selection of these EOS is justified by their compatibility with contemporary experimental and observational constraints \cite{Tews_2017}. We acknowledge, however, that this set of EOS does not include all permissible nuclear matter equations of state. 
	The cWB analysis is executed independently using a two-detector network (L1H1) in two open data intervals, designated as Time Windows TW1 and TW2, each lasting one week; Details regarding the exact durations and GPS times of these windows are listed in Table \ref{tab:Time_Window}. 
	\begin{table*}[!ht]
		\centering
		\caption{Total time of detectors network (L1H1) for the two stretches of open O3b LIGO data implemented in this study.}
		\begin{adjustbox}{max width=\textwidth}
			\begin{tabular}{| c| c@{\hspace*{1em}} | c@{\hspace*{1em}} |c@{\hspace*{1em}} | c@{\hspace*{1em}} | c@{\hspace*{1em}} |c@{\hspace*{1em}}}
				\hline
				\hline
				Time Window  & TW1 & TW1 (GPS time) & TW2 & TW2 (GPS time) \\
				\hline
				\hline
				Initial time & $2019-11-03$T$00:00:01$ & 1256774419 & 
				$2019-11-17$T$00:00:01$ & 1257465617 \\
				Final time & $2019-11-10$T$23:59:59$ & 1257984019 & $2019-11-24$T$23:59:59$ & 1258675217 \\
				\hline
				\hline
			\end{tabular}
		\end{adjustbox}
		\label{tab:Time_Window}
	\end{table*}
	Each cWB analysis generates datasets of identified events aimed at training and testing a CNN for multiclass classification designed to distinguish among five EOS variations in the \textit{Chimera} E-series model, based on their estimated HFF slopes reported in \cite{Murphy_Casallas_2024}. 
	We previously reported the DNN model's performance in estimating the HFF slope within TW1 in Table \ref{tab:HFF_TW1}; a detailed explanation can be found in \cite{Murphy_Casallas_2024}. For the study presented in this paper, we extended our analysis to the second time window (TW2), whose results in estimating the HFF slope are presented in Table \ref{tab:HFF_TW2}.
	\\ \\
	Figure \ref{fig:scatter} illustrates how the EOS influences the HFF slope estimation; This figure displays the range of variability of the estimated HFF slopes for the CCSN GW signals considered in this study.
	The gray band represents the estimated HFF slopes for the SFHo and SFHx models, while purple and violet stripes denote the DD2, FSUgold, and IUSFU models (all derived using the methodology in \cite{Murphy_Casallas_2024}) within the time window TW1. 
	For a comparative analysis, we introduce the range of variability in the HFF slope estimation derived from the TW2 dataset, keeping the same color code as described before.
	This visualization clarifies the fact that, at 1 kpc, the five EOS models, analyzed with LVK noise, can be categorized into two main, and distinct, groups based on their HFF estimated slopes. However, as the distance increases, the distinction between these groups progressively diminishes, as is evident in the plots for 5 kpc and 10 kpc, where the groups progressively merge and become indistinguishable. 
	The variability of the estimated HFF slope within the TW2 reveal a more marked separation between the two groups, compared to the TW1. This fact can be attributed to the estimated slope of HHF of the FSUgold model in the TW2 dataset, which exhibits the largest variation ($s-\hat{\bar{s}}=67$ Hz~s$^{-1}$) accounting for the difference between the slope in absence of noise, $s$ and the mean estimated slope in TW2, $\hat{\bar{s}}$, the highest variation in the estimation of the slope of HFF in the TW1 and TW2 datasets.
	\begin{table*}[!ht]
		\begin{center}
			\begin{adjustbox}{max width=\textwidth}
				\begin{tabular}{ l c c c c c c c c c c c c c } 
					\hline
					\hline
					& & \multicolumn{3}{c}{1 kpc} & \multicolumn{3}{c}{5 kpc} & \multicolumn{3}{c}{10 kpc} \\
					\cmidrule(lr){3-5}\cmidrule(lr){6-8}\cmidrule(lr){9-11}
					EOS & $s$ & $\hat{\bar{s}}$ & STD & RMSE & $\hat{\bar{s}}$ & STD & RMSE & $\hat{\bar{s}}$ & STD & RMSE\\
					& [Hz~s$^{-1}$] & [Hz~s$^{-1}$] & [Hz~s$^{-1}$] & [Hz~s$^{-1}$] & [Hz~s$^{-1}$] & [Hz~s$^{-1}$] & [Hz~s$^{-1}$] & [Hz~s$^{-1}$] & [Hz~s$^{-1}$] & [Hz~s$^{-1}$]\\
					\hline
					DD2 & 1398 & 1402 & 103.54 & 301.72 & 1742 & 293.13 & 223.11 & 2377 & 765.30 & 635.93\\ 
					
					FSUGold & 1665 & 1610 & 126.93 & 228.84 & 2002 & 258.56 & 332.91 & 2544 & 678.70 & 468.19\\ 
					
					IUFSU & 1502 & 1566 & 100.66 & 229.11 & 1812 & 201.69 & 154.25 & 2404 & 555.41 & 526.20\\ 
					
					SFHo  & 2131 & 2092 & 52.63 & 249.39 & 2501 & 313.45 & 256.74 & 2670 & 401.92 & 486.82\\ 
					
					SFHx & 2000 & 2003 & 60.43 & 159.30 & 2311 & 298.12 & 106.10 & 2687 & 368.12 & 416.60\\ 
					\hline
				\end{tabular}
			\end{adjustbox}
			\caption{Results of the HFF slope estimation in real interferometric LVK noise using the methodology described in \cite{Murphy_Casallas_2024} (refer to Section V, Table II). The analysis was conducted for Galactic distances of 1 kpc, 5 kpc, and 10 kpc, focusing on the Time Window 1 (TW1) data. The second column of the table reports the HFF slope estimation for each \textit{Chimera} E-series signal in absence of noise. For each specified detection distance, we provide the mean estimated slope, $\hat{\bar{s}}$, along with its corresponding standard deviation (STD) and root-mean-square error (RMSE). These metrics quantify the accuracy and precision of the HFF slope estimations in the presence of realistic LVK detector noise.}
			\label{tab:HFF_TW1}
		\end{center}
	\end{table*}
	\begin{table*}[!ht]
		\begin{center}
			\begin{adjustbox}{max width=\textwidth}
				\begin{tabular}{ l c c c c c c c c c c c c c } 
					\hline
					\hline
					& & \multicolumn{3}{c}{1 kpc} & \multicolumn{3}{c}{5 kpc} & \multicolumn{3}{c}{10 kpc} \\
					\cmidrule(lr){3-5}\cmidrule(lr){6-8}\cmidrule(lr){9-11}
					EOS & $s$ & $\hat{\bar{s}}$ & STD & RMSE & $\hat{\bar{s}}$ & STD & RMSE & $\hat{\bar{s}}$ & STD & RMSE\\
					& [Hz~s$^{-1}$] & [Hz~s$^{-1}$] & [Hz~s$^{-1}$] & [Hz~s$^{-1}$] & [Hz~s$^{-1}$] & [Hz~s$^{-1}$] & [Hz~s$^{-1}$] & [Hz~s$^{-1}$] & [Hz~s$^{-1}$] & [Hz~s$^{-1}$]\\
					\hline
					DD2 & 1381 & 1394 & 113.87 & 334.08 & 1797 & 313.23 & 276.99 & 2412 & 835.15 & 697.87\\ 
					
					FSUGold & 1653 & 1586 & 112.83 & 222.43 & 2031 & 277.31 & 350.01 & 2587 & 728.10 & 496.21\\ 
					
					IUFSU & 1543 & 1565 & 110.31 & 211.65 & 1895 & 265.91 & 185.34 & 2512 & 598.35 & 586.13\\ 
					
					SFHo  & 2147 & 2098 & 61.01 & 200.32 & 2567 & 365.76 & 232.23 & 2688 & 387.76 & 512.66\\ 
					
					SFHx & 1988 & 2018 & 56.97 & 176.14 & 2365 & 300.55 & 122.18 & 2695 & 400.07 & 445.44\\ 
					\hline
				\end{tabular}
			\end{adjustbox}
			\caption{Results of the HFF slope estimation using the methodology from \cite{Murphy_Casallas_2024}, applied to real interferometric noise data collected during the TW2, which was taken a week after TW1 within the LVK O3b dataset. 
				The HFF estimation for TW2 shows that the estimated slopes differ by less than 5\% compared to those from TW1, and the associated standard deviation (STD) and root-mean-square error (RMSE) values do not exceed 7\% of their TW1 counterparts. 
				The data provided in this table also indicates that the FSUgold model manifests a significant variation with respect to the slope in the noiseless case $5.7\%$. Comparatively this variation is the highest over the TW1 and TW2 datasets, corresponding to ($s-\hat{\bar{s}}=67$ Hz~s$^{-1}$). Such deviation from the noiseless case can be used to explain the clearer separation, at 1 kpc, between the two EOS groups illustrated in the bottom panel Figure \ref{fig:scatter}.}
			\label{tab:HFF_TW2}
		\end{center}
	\end{table*}
	%
	\section{HFF evolution in time and its relation with parameters of the PNS source}
	\label{sec1:physics}
	Pajkos et al. (2019) \cite{Pajkos_2019} conducted a comprehensive investigation into GW emission from CCSNe, focusing on the influence of progenitor mass and rotation in the absence of noise.
	In the rest of this section, we use the term progenitor mass to describe the mass of a star just before it explodes as a supernova. This differs, in general, from the ZAMS mass, as significant mass could be lost in the pre-explosion evolution.
	Their studies utilized 15 two-dimensional (2D) neutrino radiation-hydrodynamics simulations, spanning the critical period from initial stellar collapse to approximately 300 milliseconds after core bounce. The research systematically explored the impact of progenitor characteristics by varying four progenitor masses, specifically 12, 20, 40, and 60 $M_{\odot}$, alongside four core rotation rates ranging from 0 to 3 $rad s^{-1}$. The study specifically investigated how these progenitor properties influence the GW emission during the post-bounce, pre-explosion, and accretion phases.
	They concluded that the overall duration of the major GW emission phase was found to vary with the specific exploding progenitor, and a strong correlation exists between the total GW energy radiated and the progenitor's compactness. They also identified a critical role for centrifugal support in shaping the CCSN GW signal. They observed that PNSs exhibiting larger radii are characterized by a strong degree of centrifugal support. A distinct observation is that increasing the rate of rotation in the core essentially flattens the slope of the HFF. This spectral flattening is a direct consequence of the rotational effects that suppress high-frequency turbulent activity, leading to a redistribution of GW power towards lower frequencies.
	\\ \\
	An alternative approach to characterize the HFF signal component present in the GW emission from CCSNe has been systematically explored in the studies of \cite{Andresen_2017} and Murphy \cite{Murphy_2025}. Andresen et al. (2017) provided GW signal predictions for four models derived from three nonrotating progenitors with masses of 11.2, 20, and 27 $M_{\odot}$, using 3D multigroup neutrino hydrodynamics simulations of CCSNe. They observed that the HFF component primarily originates from aspherical mass movements within and adjacent to the overshooting region of PNS convection in 3D, whereas in two-dimensional (2D) simulations, it is related to mass motion near the gain radius. This suggests that quasi-oscillatory mass motions at high frequencies are instigated exclusively by PNS convection in 3D, even during the pre-explosion phase, while in 2D, the standing accretion shock instability (SASI) and convection in the gain region predominate.
	More recently, Murphy et al. (2025) proposed a study aimed at characterizing both the source and the emission processes of CCSN GWs, emphasizing the spatial decomposition analysis of the HFF GW emission from PNS based on 3D CCSNe simulations starting from a 15~$M{\odot}$ progenitor and a zero-metallicity 25~$M{\odot}$ progenitor. 
	Their results indicate that the GW emission is initially dominated by the surface layers of the PNS due accretion onto the PNS surface as the explosion develops. Subsequently, as the explosion progresses and accretion decreases, the GW emission becomes dominated by the regions of sustained Ledoux convection and convective overshoot deep within the PNS.
	\\ \\
	Morozova et al. (2018)\cite{Morozova_2018} conducted a foundational study on GWs from a set of two-dimensional multi-group neutrino radiation hydrodynamic simulations of CCSNe. Their primary objective was to systematize the current knowledge regarding the post-bounce CCSN GW signal and to identify ``templatable features" that could be effectively utilized by ground-based laser interferometers for detection and analysis. The study specifically compared calculated PNS oscillation frequencies from a modal analysis directly with the GW signals produced from their 2D CCSN simulations, focusing on progenitor masses of 10, 13, and 19 $M_{\odot}$. They employed a linear perturbation analysis of the angle-averaged PNS profile, demonstrating its accuracy in reproducing the dominant GW signal components.   
	The authors also reported notable quantitative differences in GW amplitudes, total GW energy emitted, and the frequencies of PNS quasi-normal-mode oscillations when comparing results across the three different EOSs they considered and for each progenitor mass. Their collection of models highlighted the strong dependence of the CCSN GW HFFs on the progenitor mass, EOS, many-body corrections to neutrino opacity, and rotation.
	\\ \\
	A recent study specifically examining the relationship between the HFF slope and progenitor properties, with a focus on varying the EOS for the \textit{Chimera} E-series, has been published by \cite{Murphy_Casallas_2024}. 
	In this investigation, the authors present how the EOS impacts the estimation of the HFF slope in the presence of LVK noise at different Galactic distances when varying the EOS without altering progenitor parameters. 
	The study reveals that while parameters like rotation rate and mass influence the HFF slope, their simultaneous variation creates more complex effects. Specifically, when the internal structure and rotation rate vary together, the reduction in the HFF slope is less pronounced than when the internal structure remains unchanged. This finding is consistent with Wang and Pan [56]. Similarly, Jardine et al. [57] show that the presence of strong magnetic fields also affects these dynamics.
	\\ \\
	Studies have shown that the CCSN GW signal is strongly influenced by its progenitor’s core structure, the EOS, and the neutrino interaction properties responsible for PNS cooling \cite{Powell_2022, Sotani_2021, Torres_Forn__2019}. These factors play a leading role in setting the slope of the GW signal's HFF. To analyze these signals, universal relations have been investigated to describe the evolution of the GW peak frequency as either an f- or g-mode \cite{M_ller_2013}. Bizouard et al. (2021) \cite{Bizouard_2021} and Bruel et al. (2023) \cite{Bruel_2023} used these relations to perform parameter estimation in Gaussian simulated noise, discussing the accuracy of the parameterization regardless of the specific EOS considered. Further work by O'Connor et al. (2013) \cite{Ott_2013} explored the parameters that describe the PNS's post-bounce evolution and its relationship with progenitor core properties, specifically highlighting the non-trivial link between the progenitor's mass and the core's compactness parameter.
	\section{Methodology}
	\label{sec:Methodology}
	%
	The current study combines the initial HFF slope estimation technique \cite{Casallas_2023} with an image-based pattern recognition model, specifically a CNN \cite{CNN1, CNN2} as follows:
	(i) \textbf{Generation of cWB events and processing of likelihood time-frequency maps}: The cWB algorithm is used to generate likelihood time-frequency maps, $L$, for events identified through event production analysis. The configuration parameters for cWB align with those used in \cite{Casallas_2023, Murphy_Casallas_2024}. These likelihood time-frequency maps are further analyzed to extract the physical information, as elaborated in Sections II-B, II-C of \cite{Casallas_2023}, IV-A and IV-B of \cite{Murphy_Casallas_2024}.
	(ii) \textbf{DNN model}: The primary instrument for estimating the initial slope of the HFF is the DNN, as elaborated in \cite{Casallas_2023} and more recently in \cite{Murphy_Casallas_2024}. 
	(iii) \textbf{A multiclassifier CNN model}: A CNN is employed to distinguish between the EOS as in \cite{Murphy_Casallas_2024}. The CNN model receives images as input, specifically processed likelihood time-frequency maps, $L$, and produces multiclass labels reflecting the estimated slopes related to each EOS. Our classification algorithm includes: 
	($\textbf{A}$) a CNN model configured for multi-classification, which aids in distinguishing the estimated slopes at three Galactic distances; ($\textbf{B}$) 
	independently trained and tested CNN models, using data from TW1 and TW2. To assess the generalizability and robustness of the multiclassifier CNN model it was trained using the TW1 dataset and then tested on an TW2 dataset. This approach ensures a more robust evaluation of the model's performance and operation \cite{CNN1}. Finally, ($\textbf{D}$) a set of performance metrics \cite{CNN2, CNN_GW_1, CNN_GW_2} to evaluate the model's accuracy in classifying the HFF slopes for each category are introduced, this includes: accuracy, the micro and macro average one vs rest (OvR) area under the curve (AUC), and the receiver operating characteristics curve (ROC).
	\subsection{Generation of cWB Events and Processing of Likelihood Maps}\label{subsec:cWB}
	The CCSN GW signals were injected into detector strain data of the second half of the third observing run O3b, in two different time windows TW1 and TW2, at intervals of 50 seconds, at 1 kpc, 5 kpc, and 10 kpc. Every signal is modeled to propagate in an equatorial orientation relative to the detector. Our findings cover the Galactic distances examined at this equatorial alignment. However, different orientations can be realized by modifying the factor $1/r$ to incorporate multiplication by the cosine of the angle of orientation $\theta$, as explained in \cite{Creighton}. The number of detections varied with Galactic distances and cWB operational settings. The detections recorded during TW1 and TW2 are presented in Table \ref{tab:cWB_detections}. 
	\begin{table*}
		\centering
		\caption{Total number of detections obtained from the cWB event production for the five EOS; DD2, FSUgold, IUSFU, SFHo and SFHx for the two stretches of open O3b LIGO data implemented in this study.}
		\begin{adjustbox}{max width=\textwidth}
			\begin{tabular}{| c| c@{\hspace*{1em}} | c@{\hspace*{1em}} |c@{\hspace*{1em}} | c@{\hspace*{1em}} | c@{\hspace*{1em}} |c@{\hspace*{1em}}| c@{\hspace*{1em}} | c@{\hspace*{1em}}}
				\hline
				\hline
				EOS  & TW1 1kpc & TW1 5kpc & TW1 10kpc & TW2 (1kpc) & TW2 (5kpc) & TW2 (10kpc) \\
				\hline
				\hline
				DD2 & 1002 & 913 & 613 & 1044 & 922 & 631 \\
				FSUgold & 987 & 801 & 601 & 966 & 754 & 567 \\
				IUSFU & 1034 & 987 & 687 & 1066 & 1000 & 702 \\
				SFHo & 3787 & 3215 & 2215 & 3821 & 3299 & 2199 \\
				SFHx & 3211 & 3034 & 2034 & 3266 & 3087 & 2021 \\
				\hline
				\hline
				Total  & 10.021 & 8.950 & 6.150 & 10.163 & 9.062 & 6.120 \\
				\hline
				\hline
			\end{tabular}
		\end{adjustbox}
		\label{tab:cWB_detections}
	\end{table*}
	Analysis of detected events reveals a significant class imbalance \cite{ML1, ML2} across both TW1 and TW2 datasets. Specifically, the SFHo and SFHx classes represent the most dominant signals, a fact in agreement with what was reported in \cite{Murphy_Casallas_2024}, consistently exhibiting the largest number of detections throughout both time windows. In contrast, the DD2, FSUgold and IUSFU classes account for approximately 30\% of the detections compared to SFHo and SFHx. 
	\\ \\
	The distribution of detections indicates that the datasets implemented for estimating the HFF from CCSN GW signals present a class imbalance persisting in TW1 and TW2. The distribution of EOS classes reveals that DD2, FSUgold, and IUSFU classes are  underrepresented, compared to the predominant SFHo and SFHx classes. 
	This imbalance stems from the inherent physics of how these EOSs influence GW emission and their subsequent detectability. Specifically, the underrepresented EOSs likely yield GW signals that are intrinsically weaker or possess HFF characteristics (e.g., lower amplitude, less distinct time evolution) that place their energy predominantly in frequency bands with higher LVK noise. Following the results of \cite{Landfield_thesis} softer EOS models, such as SFHo and SFHx, describe more compressible matter. This produces a PNS with a smaller radius and higher central density. 
	The compact structure supports higher-frequency oscillation modes, especially the surface g-mode, which dominates the late post-bounce GW emission. As a result, the GW spectrum for soft EOSs is shifted toward the kilohertz range, with sharper peaks that are well matched to power-based burst search algorithms such as cWB.
	In contrast, stiffer EOS models like DD2 and FSUGold describe less compressible matter. 
	The resulting PNS has a larger radius and lower central density, which shifts its oscillation frequencies downward. The GW emission from these models is typically weaker at high frequencies and instead dominated by longer-duration, lower-amplitude contributions from neutrino-driven convection and the standing accretion shock instability (SASI). While energetically significant, these signals distribute their power over longer timescales and at lower frequencies, making them harder to distinguish above the detector noise floor.
	%
	\subsection{Convolutional neural network architecture}\label{subsec:CNN}
\begin{figure*}
		\centering
		\includegraphics[width=1.0\textwidth]{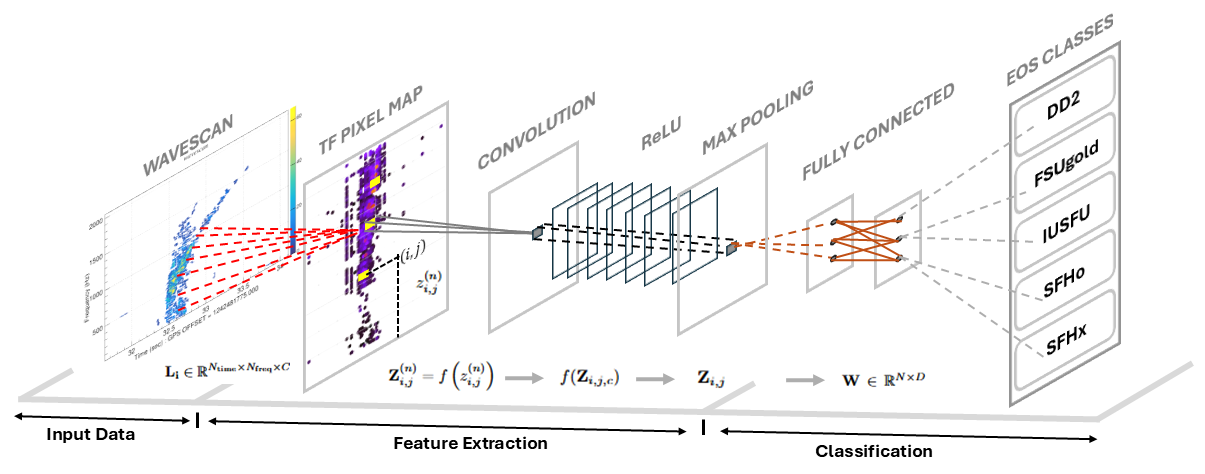}
		\caption{
			Single-stack CNN architecture implemented for the classification of the EOS at different Galactic distances within LVK noise. CNN input consists of time-frequency likelihood maps ($L_i$), which are derived from cWB event production using wavescan. Each time-frequency likelihood map encodes the CCSN GW HFF slope within its internal pixel configuration. The CNN extracts spatial features from these input images using a convolution layer (Equation \ref{E:Conv}), where a filter slides across the entire image. This is followed by a non-linear activation function, such as ReLU (Equation \ref{E:ReLU}), which allows the network to learn complex patterns. A subsequent max-pooling layer reduces the spatial dimensions, while finally a fully-connected layer merges the learned features to make a classification decision (Equation \ref{E:Feature}). For a detailed description of the CNN architecture and its performance, refer to Appendices \ref{Ap:CNN_model} and \ref{Ap:Performance}.    
		}
		\label{fig:CNN_model}
	\end{figure*}

	%
	Unlike our previous works \cite{Casallas_2023, Murphy_Casallas_2024}, which employed a Deep Neural Network (DNN) that processed a standardized vector representation of the data to estimate the HFF slope, the CNN developed here directly processes time-frequency likelihood maps (See section IV-A in \cite{Murphy_Casallas_2024} for details) as image inputs. Each image visually encodes the likelihoods maps associated with different HFF slope values for a given EOS.
	\\ \\
	For this study, we used a CNN for multi-classification because of its versatility in : (i) performing pattern recognition within images, by learning from internal features organized in grids of neurons that can be used across the entire image, piece by piece \cite{Antelis_2022, ML1}, (ii) it requires fewer parameters \cite{CNN_GW_1} to achieve learning compared to a DNN, thus enhancing efficiency \cite{ML2}, (iii) it is specifically designed to learn and generalize features directly from images provided as input \cite{CNN1, CNN2, Mitra_2024}, and (iv) it demands less pre-processing than other classification algorithms  \cite{CNN_GW_1, CNN_GW_3}. 
	In a CNN model, the input data is passed through a series of layers that are designed to extract, in a hierarchical structure, increasingly abstract features embedded within images, and then do the classification \cite{CNN_GW_2}. The basic building blocks of a CNN (see Appendix \ref{Ap:CNN_model}), unlike DNN, are convolutional layers, which use filters to extract features from input data, and pooling layers, which down-sample the output of the convolutional layers to reduce the dimensionality of the data \cite{ML1, CNN1}. After passing through several convolutional and pooling layers, the output is then transformed into a single vector and fed into a series of fully connected layers that perform classification or regression of the extracted features \cite{Mitra_2024}. 
	Convolution, responsible for executing an affine transform \cite{CNN_GW_4}, detection \cite{CNN_GW_2}, aimed at recognizing nonlinearities, and pooling \cite{CNN_GW_3}, which serves the function of downsampling, constitute the fundamental substages implemented within layers that collectively form a stack. Subsequently, this stack can be sequentially connected to additional stacks, culminating in the final stack interfacing with the classifier. 
	Figure \ref{fig:CNN_model} illustrates the operation of the single-stack CNN model designed for EOS classification. For further details regarding the internal composition of the CNN, we refer to Appendices \ref{Ap:CNN_model} and \ref{Ap:CNN_model}. 
	In this study, we employed a single-stack CNN architecture. Our primary objective is not to optimize classification efficiency through architectural variations, but rather, we aim to ascertain the viability of differentiating among the five equations of state (EOS) across three Galactic distances. 
	The CNN architecture designed for this research is detailed in Table \ref{tab:CNN model}. This CNN model contains a total of 545,589 trainable parameters.  
	\begin{table}[!ht]
		\centering
		\caption{Architectural description of the single stack CNN multi-classifier model implemented to classify the EOS from the \textit{Chimera} E-series using cWB detections.}
		\begin{adjustbox}{max width=\textwidth}
			\begin{tabular}{c| c@{\hspace*{1em}} c@{\hspace*{1em}} c@{\hspace*{1em}} c@{\hspace*{1em}} c@{\hspace*{1em}}}
				\hline
				\hline
				Layer  & Output Shape \\
				\hline
				\hline
				Conv2d\_23 (Conv2D)  & (None,$28, 28, 32$)\\
				leaky\_re\_lu\_25 (LeakyReLU) & (None,$28, 28, 32$)\\
				max\_pooling2d\_21 (MaxPooling2D) & (None,$14, 14, 32$)\\
				Conv2d\_24 (Conv2D) & (None,$14, 14, 64$)\\
				leaky\_re\_lu\_26 (LeakyReLU) & (None,$14, 14, 64$)\\
				max\_pooling2d\_22 (MaxPooling2D) & (None,$7, 7, 64$)\\
				Conv2d\_25 (Conv2D) & (None,$7, 7, 128$) \\
				leaky\_re\_lu\_27 (LeakyReLU) & (None,$7, 7, 128$)\\
				max\_pooling2d\_23 (MaxPooling2D)  & (None,$4, 4, 128$)\\
				flatten\_7 (Flatten) & (None,$2048$)\\
				dense\_14 (Dense)  & (None,$128$)\\
				leaky\_re\_lu\_28 (LeakyReLU) & (None,$128$) \\     
				dense\_15 (Dense) & (None,$5$)\\
				\hline
				\hline
			\end{tabular}
		\end{adjustbox}
		\label{tab:CNN model}
	\end{table}
	%
	\section{Results}
	\label{sec:results}
	%
	\textcolor{blue}{The effectiveness and reliability of the proposed CNN multi-class classification were first assessed using a Monte Carlo cross-validation approach on each time window (TW1 and TW2) dataset (see Appendix \ref{Ap:Performance}). In each iteration, the datasets were randomly split into training (70\%) and testing (30\%) sets, a procedure repeated 30 times to mitigate the impact of random partitioning and to rigorously evaluate the distinguishability among the five EOS classes. Building on this validation, the following section evaluates the performance of the architecture across three different studies: Studies 1 and 2 (TW1 and TW2) assess the model's performance within consistent time windows, while Study 3 (TW1--TW2) examines cross-temporal robustness by testing the TW1-trained model on the independent TW2 dataset.}
	
	\subsection{Comparative Analysis of EOS Classification}\label{sec:results_analysis}
\begin{figure*}[!ht]
		\centering
		\includegraphics[width=0.32\textwidth]{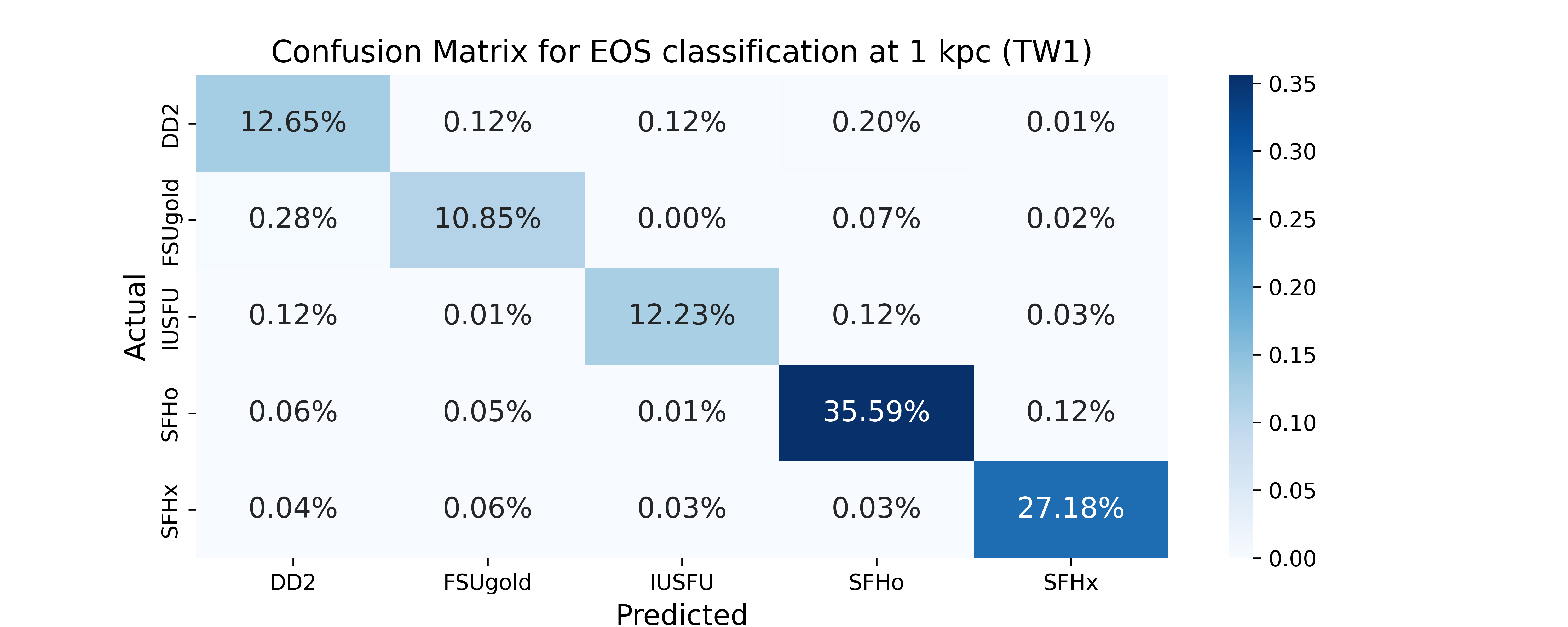}
		\includegraphics[width=0.32\textwidth]{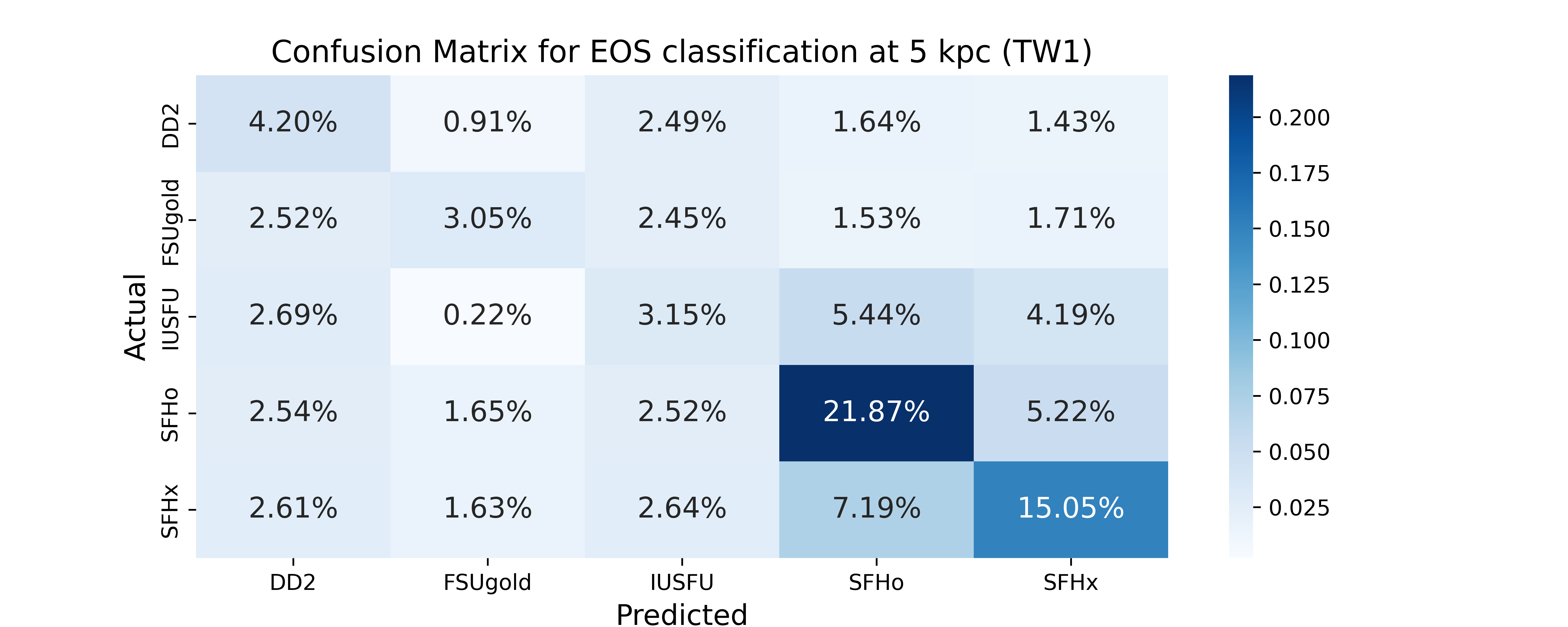}
		\includegraphics[width=0.32\textwidth]{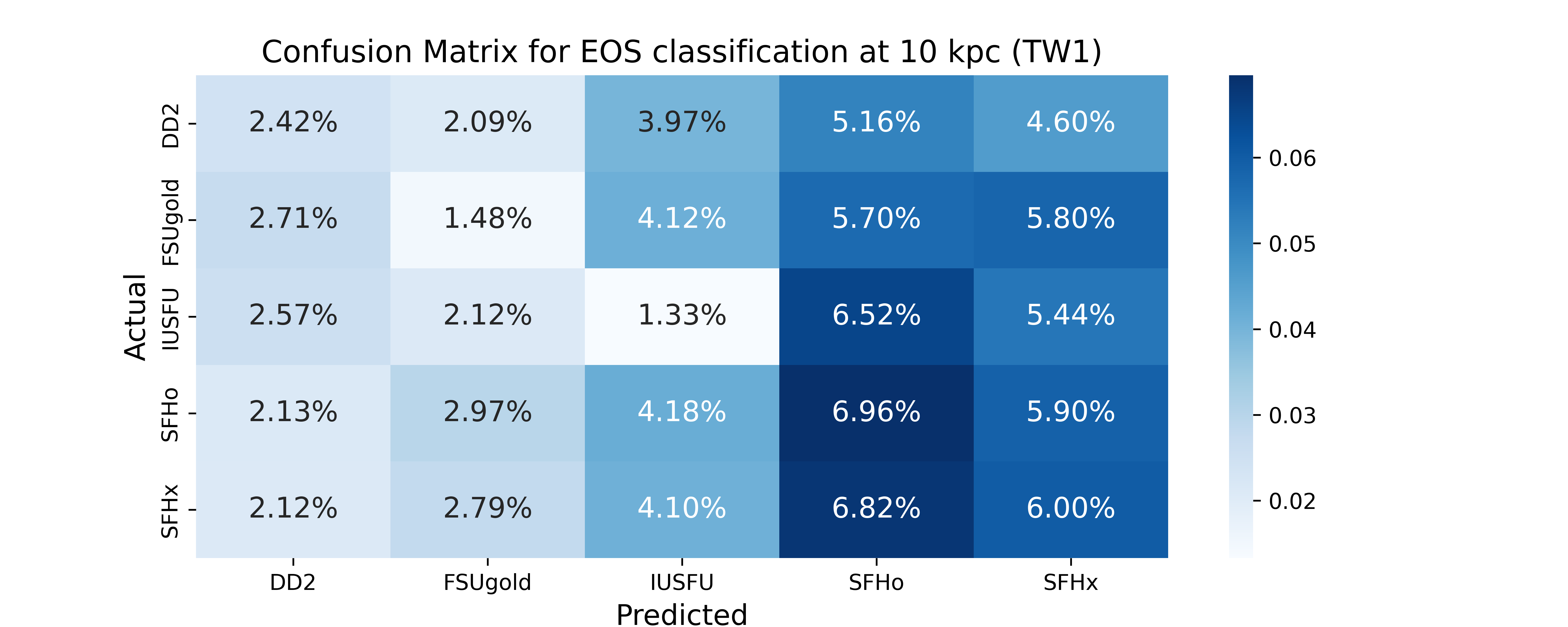} \\
		\vspace{0.2cm}
		\includegraphics[width=0.32\textwidth]{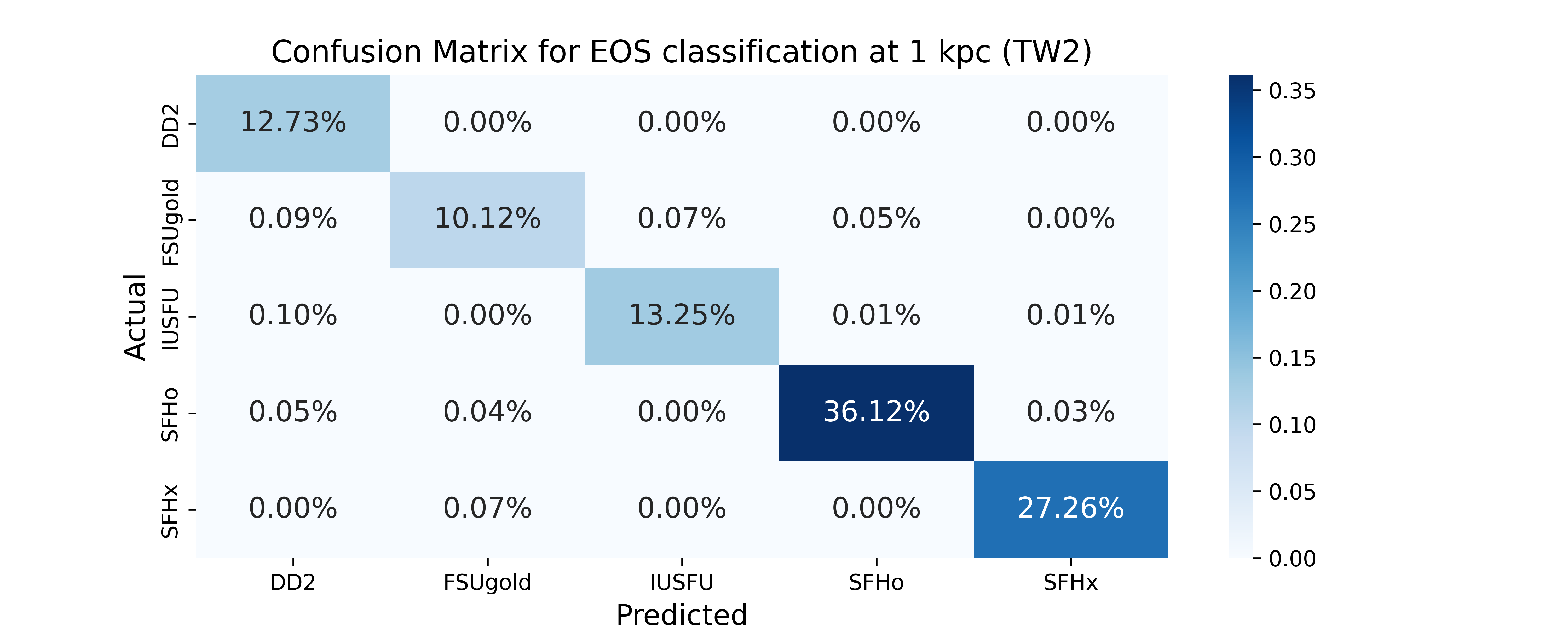}
		\includegraphics[width=0.32\textwidth]{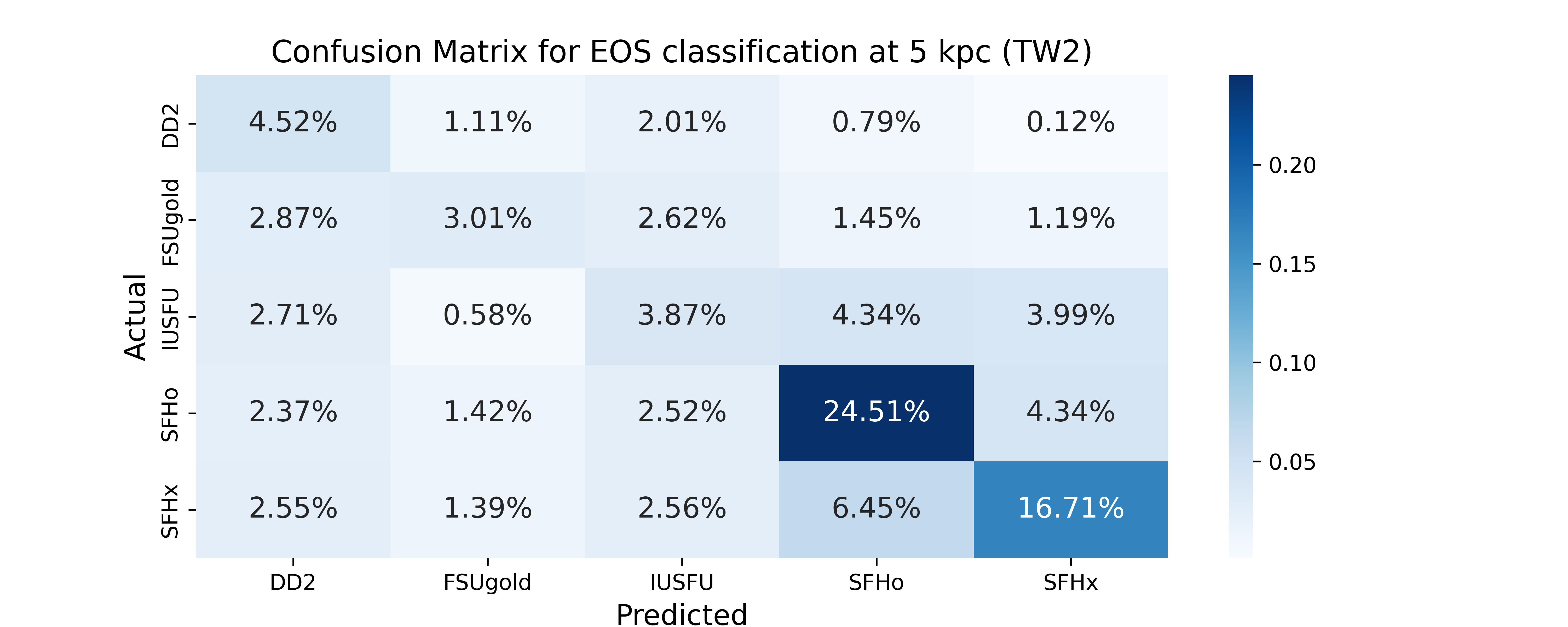}
		\includegraphics[width=0.32\textwidth]{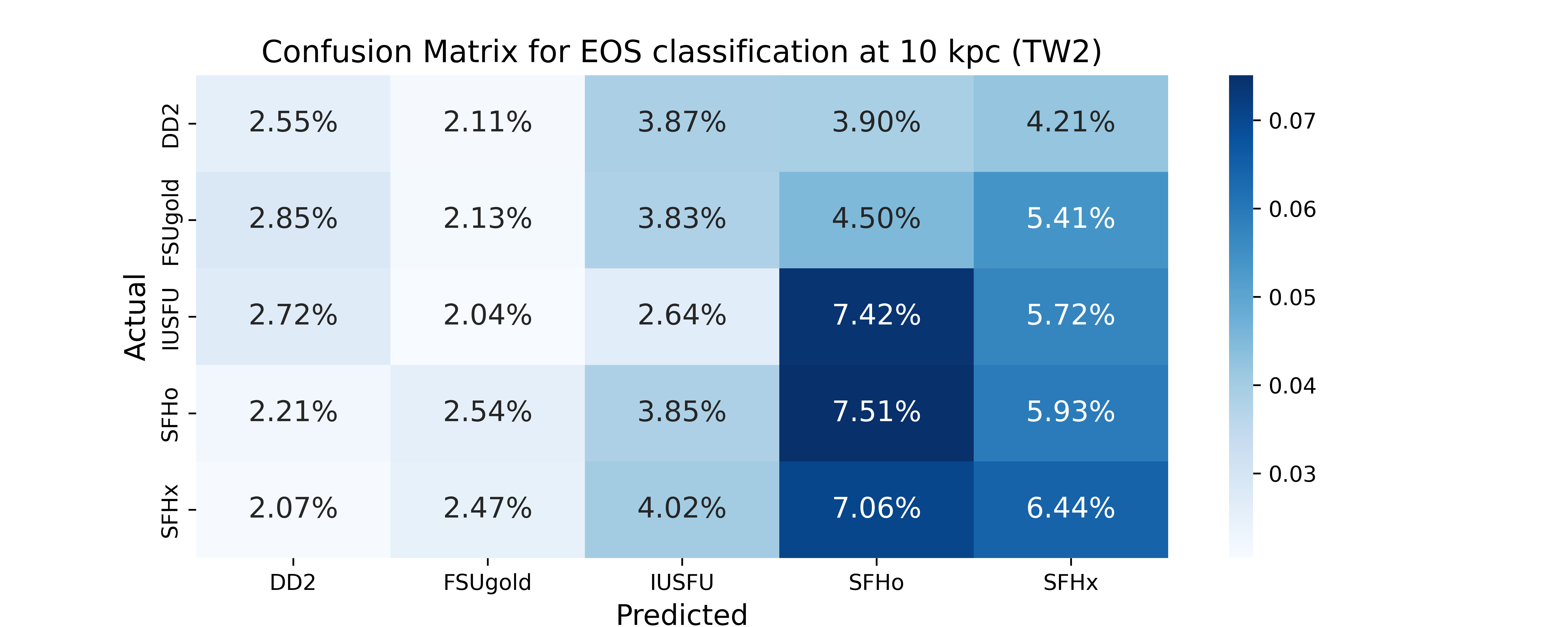} \\
		\vspace{0.2cm}
		\includegraphics[width=0.32\textwidth]{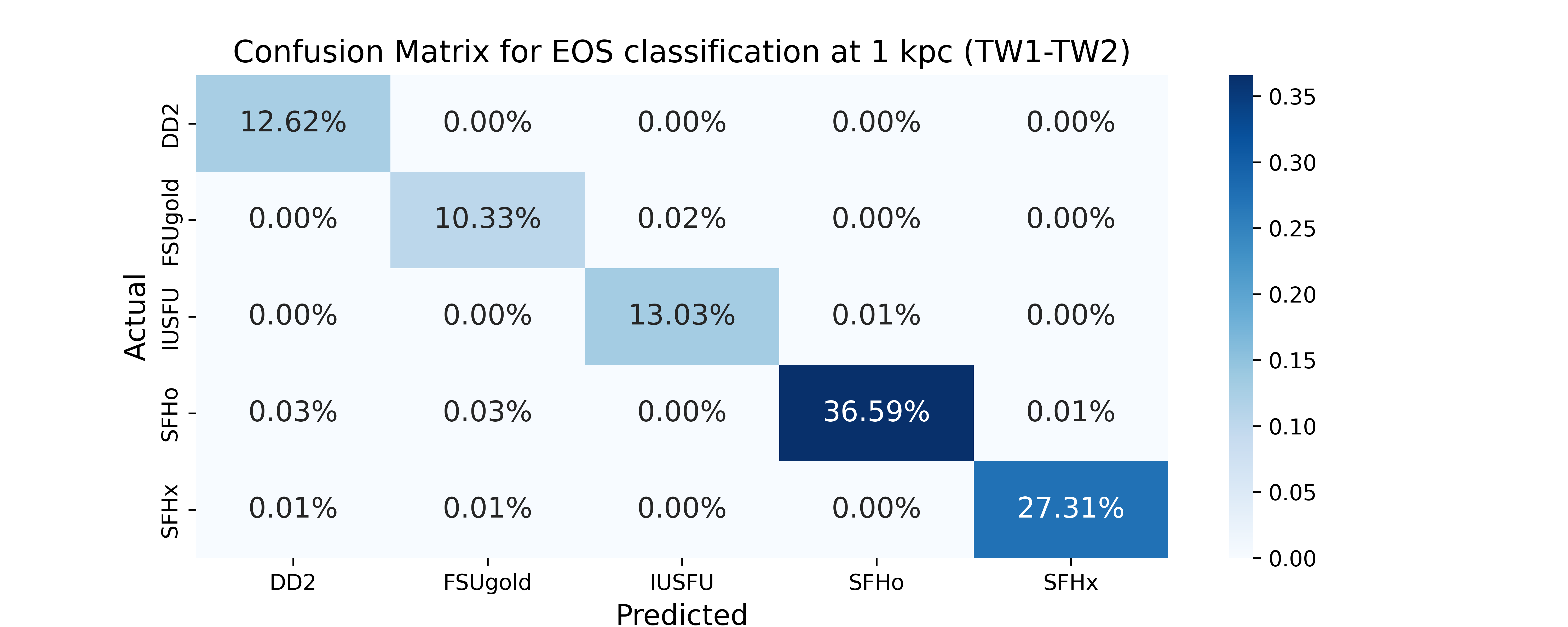}
		\includegraphics[width=0.32\textwidth]{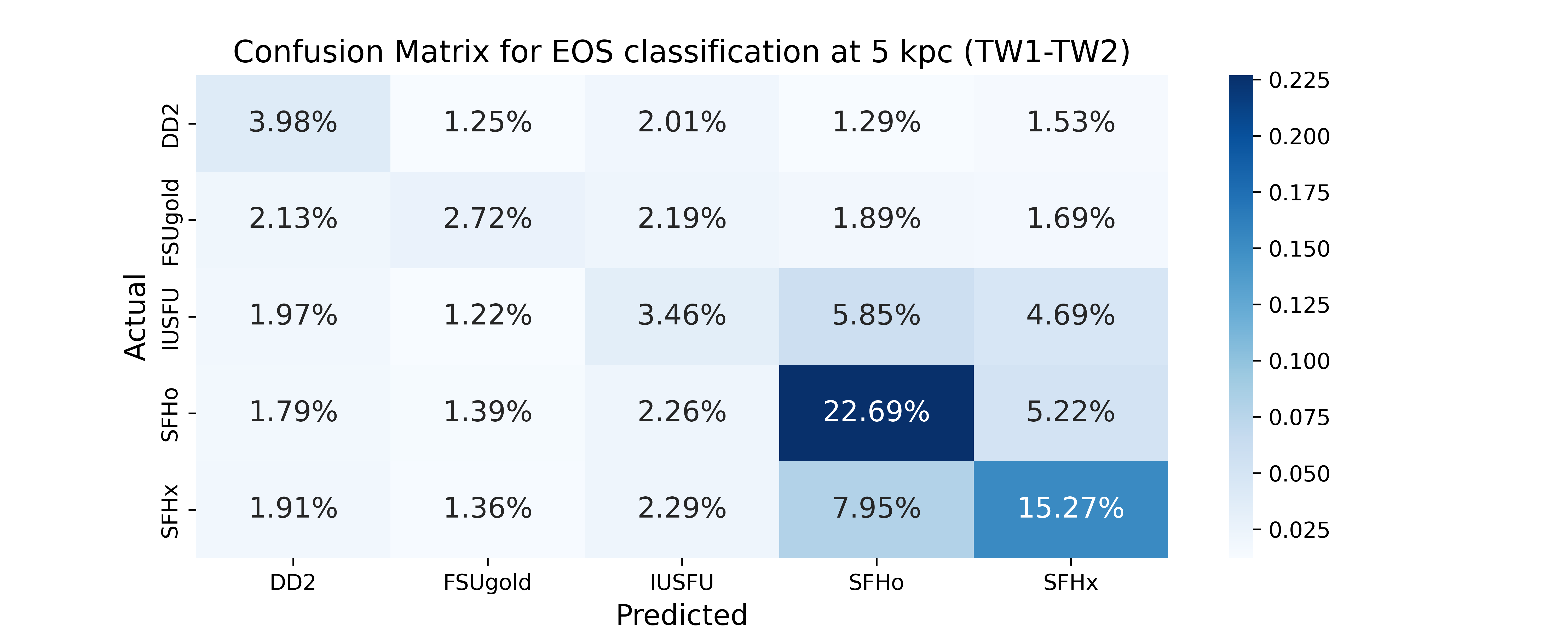}
		\includegraphics[width=0.32\textwidth]{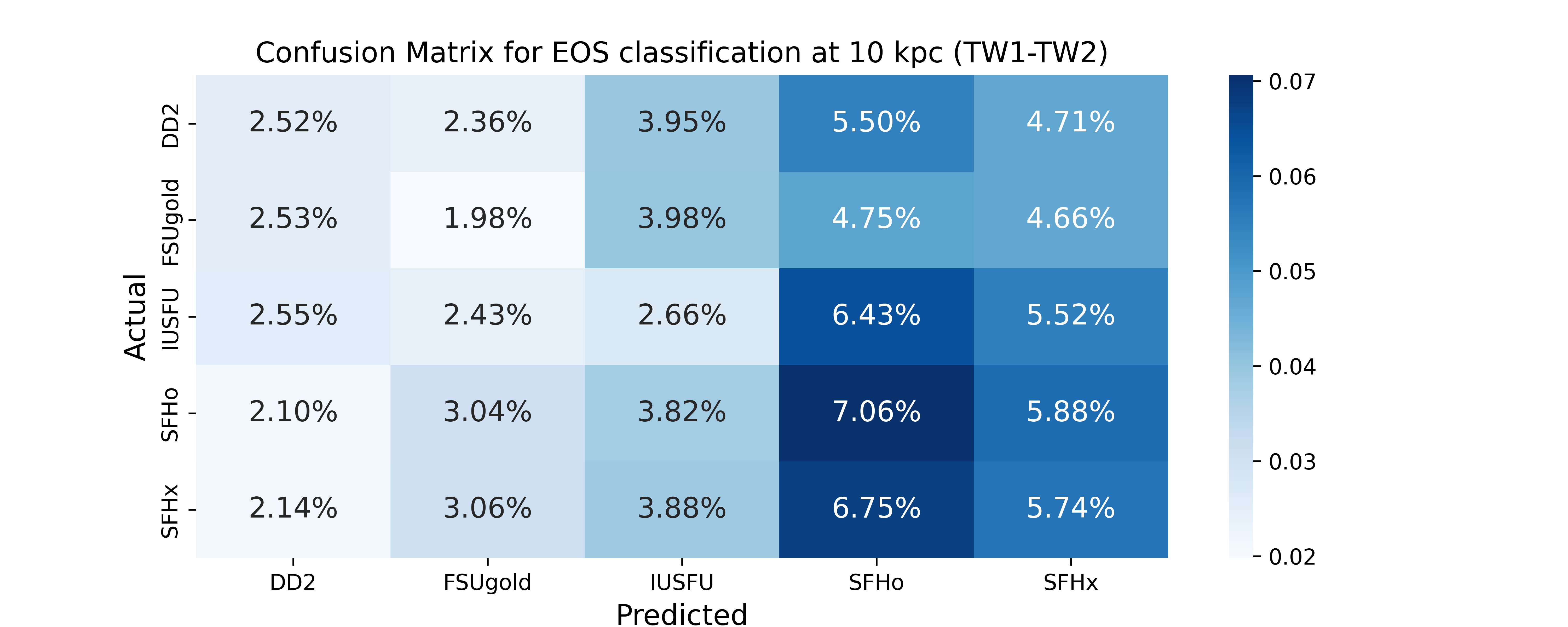}
		\caption{Confusion matrices for Study 1 (top), Study 2 (middle), and Study 3 (bottom) at distances of 1, 5, and 10 kpc (columns). At 1 kpc, the CNN model demonstrates high accuracy across all studies, successfully distinguishing between all EOS classes. As distance increases, the signal-to-noise ratio decreases, leading to a systematic degradation of the diagonal elements. The SFH-series shows the highest resilience to noise, while the IUSFU class exhibits the most significant misclassification rates at larger distances, aligning with the HFF slope deviations reported in \cite{Murphy_Casallas_2024}.}
		\label{fig:ALL_CM}
	\end{figure*}

	\textcolor{blue}{Figure \ref{fig:ALL_CM} presents the full suite of confusion matrices for all three studies across Galactic distances of 1, 5, and 10 kpc. These matrices are globally normalized, where each cell represents the fraction of total predictions for a specific (true, predicted) category. Elements along the main diagonal represent correctly classified instances, while off-diagonal elements indicate specific misclassified elements.} 
	\textcolor{blue}{As summarized in Table \ref{tab:PREDICTIONS}, the CNN model demonstrates high accuracy at 1 kpc across all considered datasets, with per-class classification rates consistently exceeding 96\%. This high performance at close range indicates that the model has successfully learned the unique HFF features associated with each EOS.
		As expected, the predictive capacity diminishes with increasing distance. At 5 kpc, a notable drop in performance occurs. While the SFHo and SFHx models remain relatively resilient, the correct classification rates for DD2, FSUgold, and IUSFU fall significantly. Notably, Study 2 (TW2) and Study 3 (TW1--TW2) exhibit slightly higher accuracy scores at 1 kpc and 5 kpc compared to Study 1. This improvement is consistent with the higher micro- and macro-average training scores obtained for the TW2-related models (refer to Tables \ref{tab:OvR_TW1} and \ref{tab:OvR_TW2}).
		At a distance of 10 kpc, the results reveal a near-total loss of classification ability for most classes, as the interferometric noise dominates the underlying signal. The misclassification rates at this distance become comparable to the correct classification rates, marking the effective limit of the current algorithm's sensitivity for the \textit{Chimera} E-series. However, the robustness demonstrated in Study 3 suggests that once trained, the CNN model remains effective against the temporal non-stationarity of the O3b data, preserving its classification logic across different observation windows.}
	\begin{table*}[!ht]
		\centering
		\caption{Per-class classification accuracy (percentage of correctly predicted samples) for the three studies across Galactic distances. Values are derived from the consolidated confusion matrices in Fig. \ref{fig:ALL_CM}.}
		\begin{adjustbox}{max width=\textwidth}
			\begin{tabular}{l | ccc | ccc | ccc}
				\hline\hline
				& \multicolumn{3}{c|}{\textbf{1 kpc}} & \multicolumn{3}{c|}{\textbf{5 kpc}} & \multicolumn{3}{c}{\textbf{10 kpc}} \\
				\textbf{Class} & \textbf{TW1} & \textbf{TW2} & \textbf{TW1-2} & \textbf{TW1} & \textbf{TW2} & \textbf{TW1-2} & \textbf{TW1} & \textbf{TW2} & \textbf{TW1-2} \\
				\hline
				DD2     & 96.19 & 98.14 & 99.68 & 28.84 & 30.09 & 33.78 & 20.25 & 20.56 & 21.83 \\
				FSUgold & 97.38 & 98.92 & 99.61 & 40.88 & 40.07 & 34.25 & 12.38 & 18.86 & 15.38 \\
				IUSFU   & 98.70 & 99.47 & 99.84 & 23.77 & 28.49 & 28.33 & 7.51  & 14.49 & 14.54 \\
				SFHo    & 99.38 & 99.83 & 99.97 & 58.05 & 65.64 & 57.19 & 22.33 & 24.71 & 23.15 \\
				SFHx    & 99.34 & 99.85 & 99.96 & 55.61 & 63.41 & 53.76 & 21.62 & 23.24 & 21.65 \\
				\hline\hline
			\end{tabular}
		\end{adjustbox}
		\label{tab:PREDICTIONS}
	\end{table*}
	\section{Dataset Balancing and SMOTE Implementation}\label{sec:smote}
\begin{figure*}[!ht]
		\centering
		\includegraphics[width=0.32\textwidth]{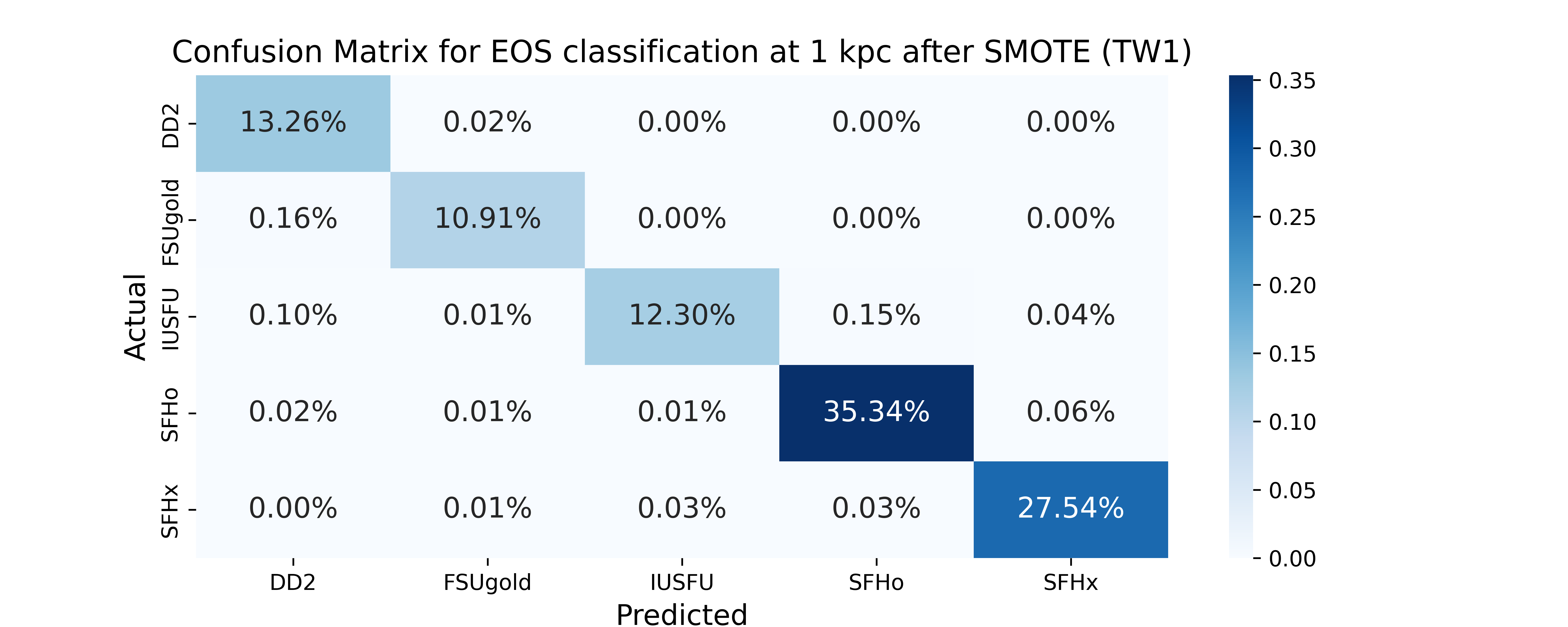}
		\includegraphics[width=0.32\textwidth]{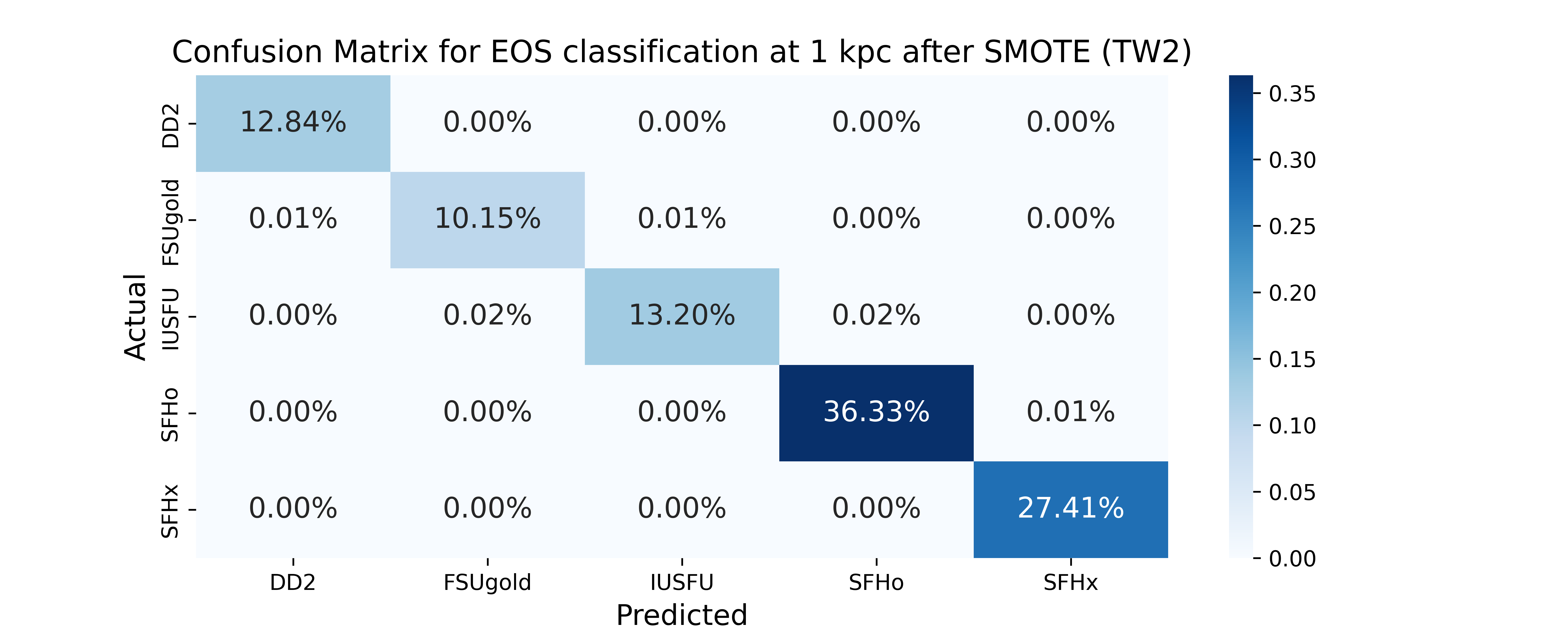}
		\includegraphics[width=0.32\textwidth]{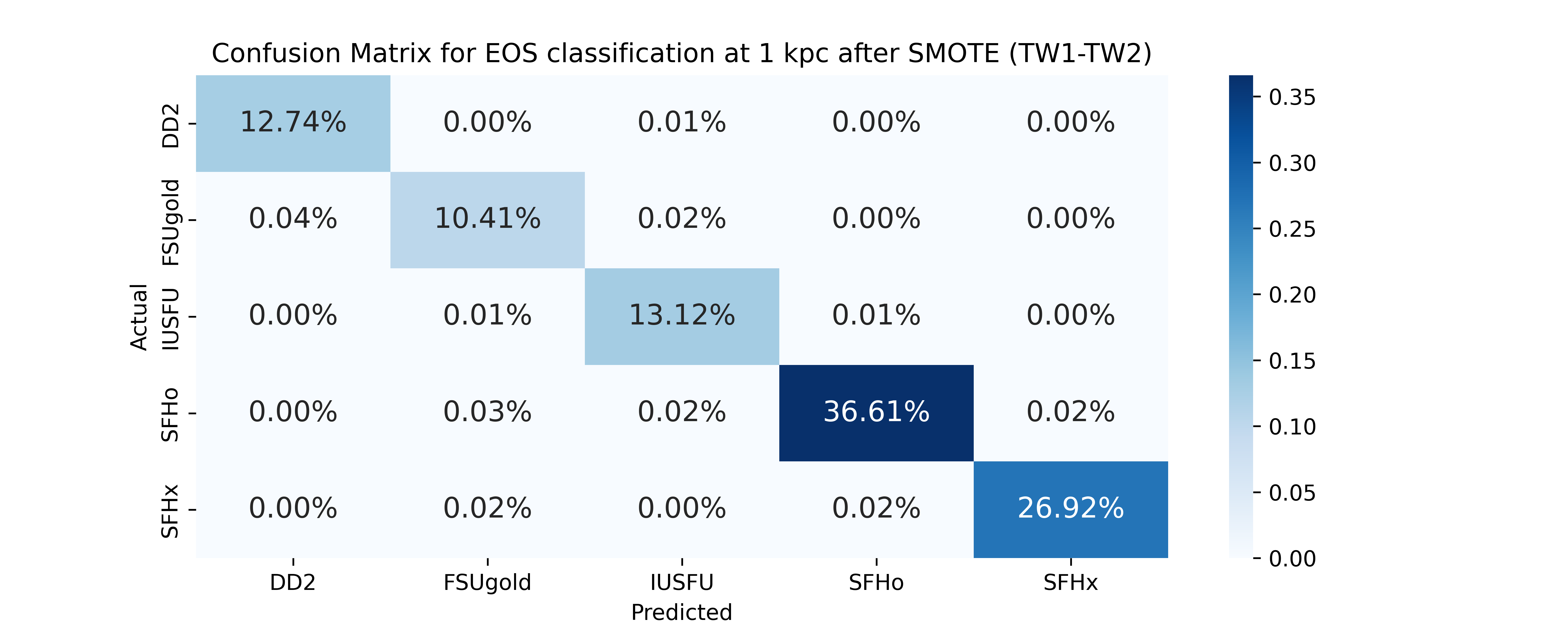} \\
		\caption{Applying the SMOTE technique to balance the datasets did not significantly alter the total prediction rates in the confusion matrices for the TW1, TW2, and TW1-TW2 datasets at 1 kpc. The fraction of total detections showed a minimal increase, less than 5\%, suggesting that SMOTE did not generate a sufficient number of synthetic samples to effectively equalize the imbalanced datasets.}
		\label{fig:SMOTE}
	\end{figure*}

	\textcolor{blue}{As previously noted in section \ref{sec:Methodology}, datasets derived from cWB-XP event production for estimating the initial slope of the HFF in CCSN GW \textit{Chimera} E-series exhibit an unbalanced configuration. Specifically, the DD2, FSUgold, and IUSFU classes contain fewer elements compared to the SFHo and SFHx classes, which have a significantly larger number of elements (see table \ref{tab:cWB_detections}). 
		To address the underrepresentation of the DD2, FSUgold, and IUSFU classes, we used the Synthetic Minority Over-sampling Technique (SMOTE). This widely used preprocessing method mitigates class imbalance by generating synthetic examples of minority classes \cite{ML1,ML2}, thus creating a more balanced dataset for testing our multi-class CNN architecture for EOS classification.
		SMOTE generates new synthetic minority instances situated between existing minority instances by means of linear interpolation for the minority class. These synthetic instances are created by randomly choosing one or more of the k-nearest neighbors for each minority class example \cite{nunes2024deeplearning, CNN1}. 
		Figures \ref{fig:SMOTE} present the confusion matrices obtained after the implementation of the SMOTE technique at 1 kpc for the TW1, TW2, and TW1-TW2 datasets.}
	\textcolor{blue}{Following the implementation of the SMOTE technique, a slight increase in the overall fraction of correctly predicted EOS classes is observed in Figure \ref{fig:SMOTE}. A comparison of the confusion matrices before and after SMOTE (Figures \ref{fig:ALL_CM}, and \ref{fig:SMOTE}) reveals that the number of synthetic samples does not significantly alter the results. 
		The data particularly indicates that, at a distance of 1 kpc, the percentage of accurate predictions rises by under 5\% with the application of SMOTE. Although slight, this increase highlights the reliability of the CNN model. Both the algorithm's effectiveness and the distribution of its predicted probabilities are stable, showing uniformity in the density of correctly classified samples by class, in all three studies, irrespective of the initial class distribution detailed in Table \ref{tab:cWB_detections}.}
	\section{Receiver operating curves for EOS identification}
\begin{figure*}
		\centering
		\includegraphics[width=0.32\textwidth]{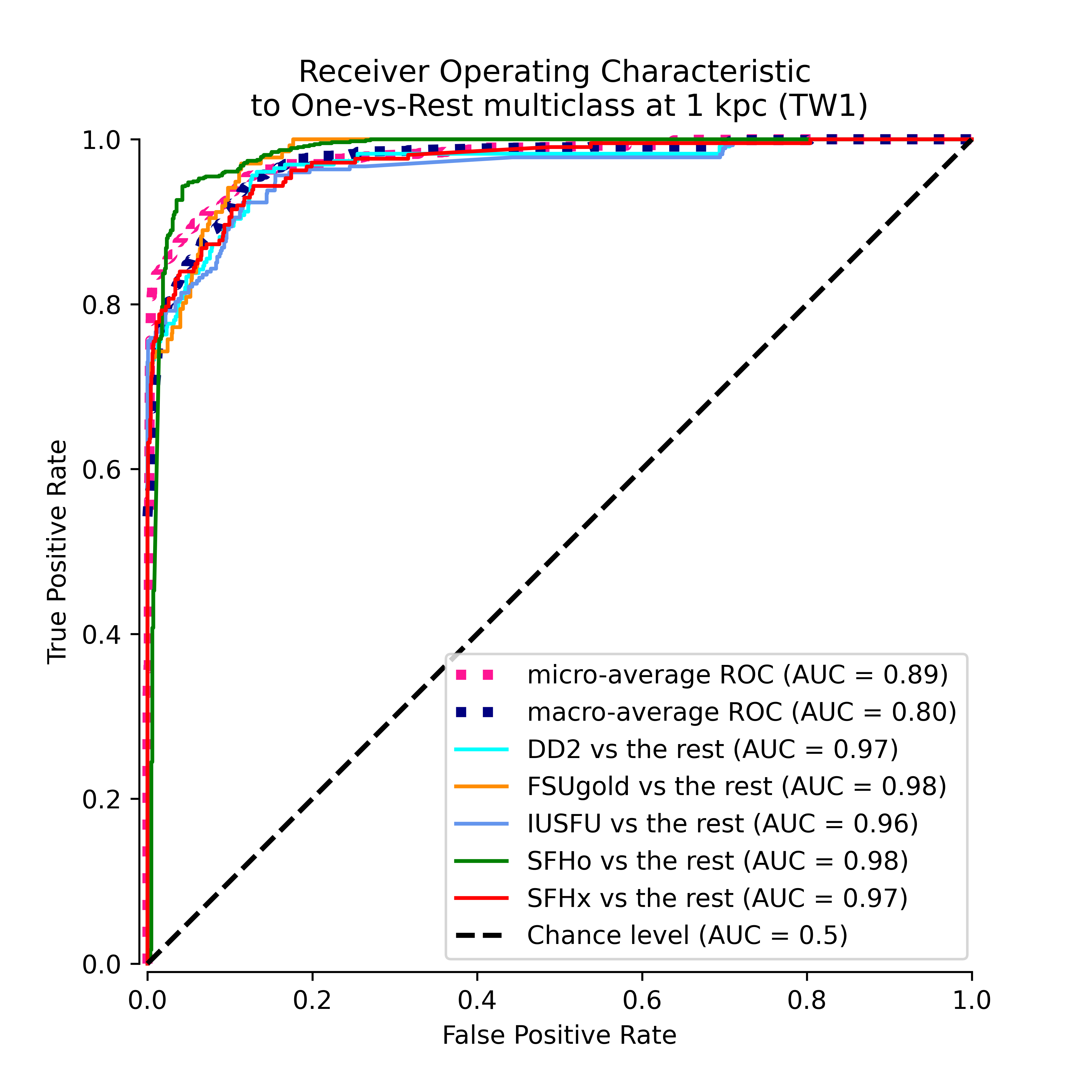}
		\includegraphics[width=0.32\textwidth]{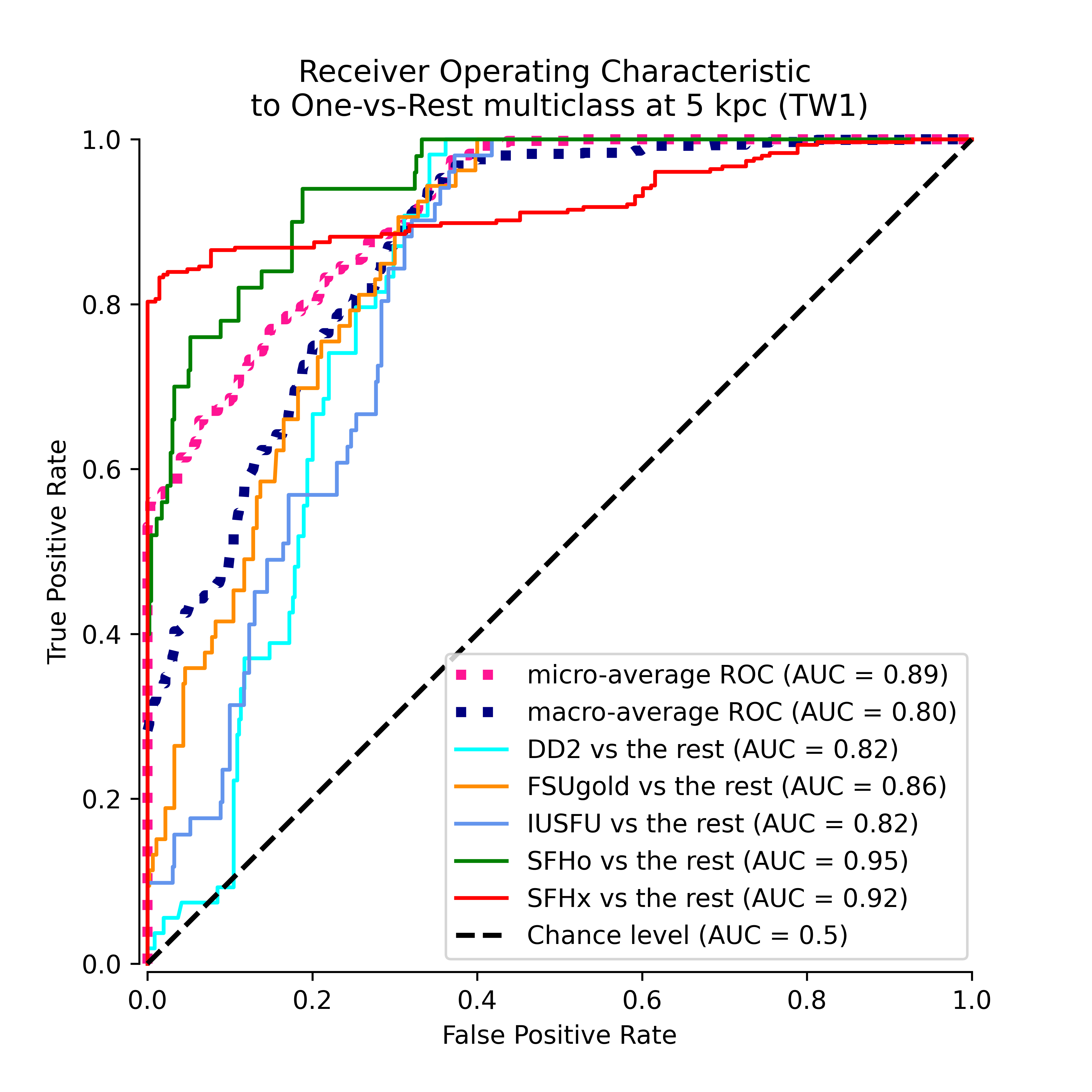}
		\includegraphics[width=0.32\textwidth]{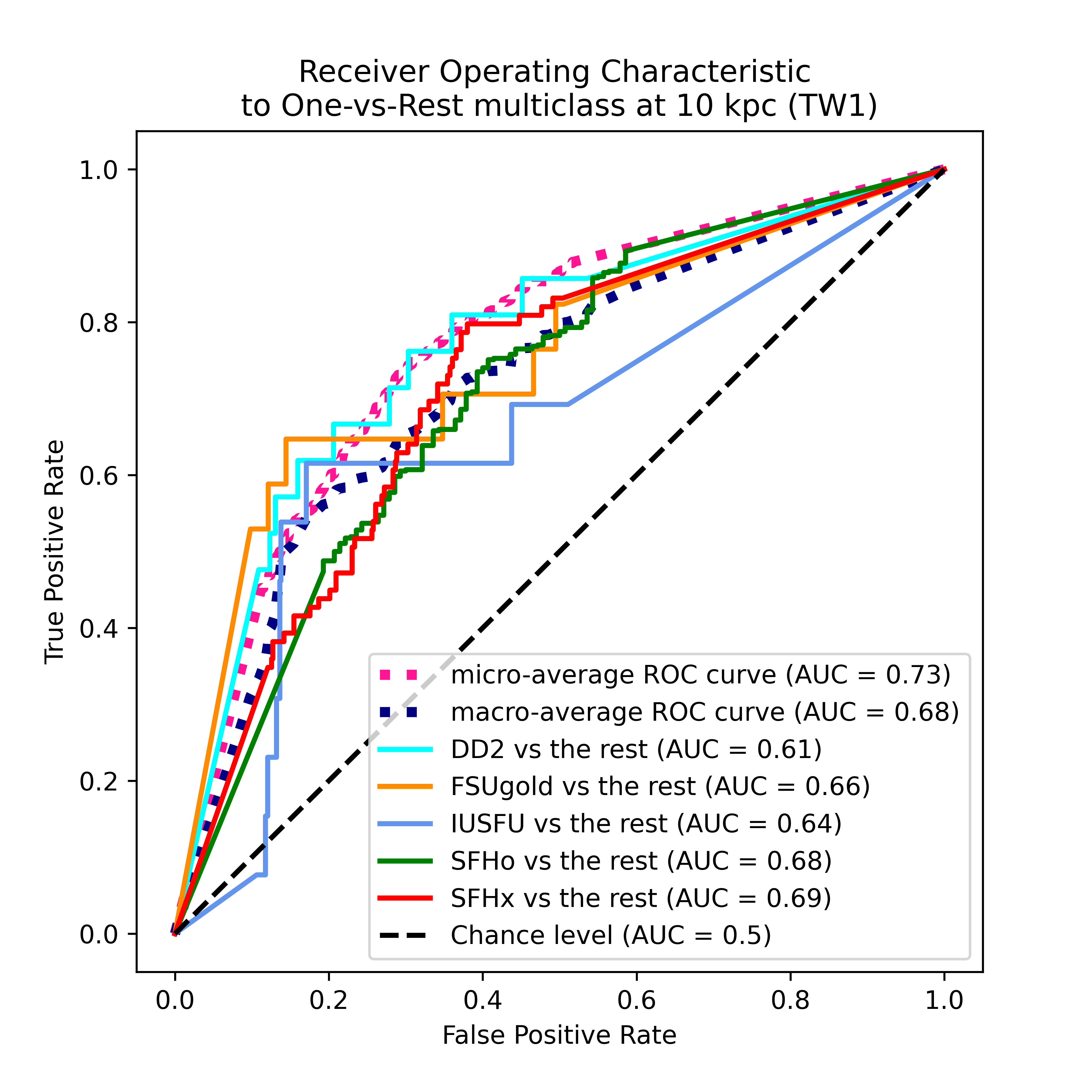}
		\includegraphics[width=0.32\textwidth]{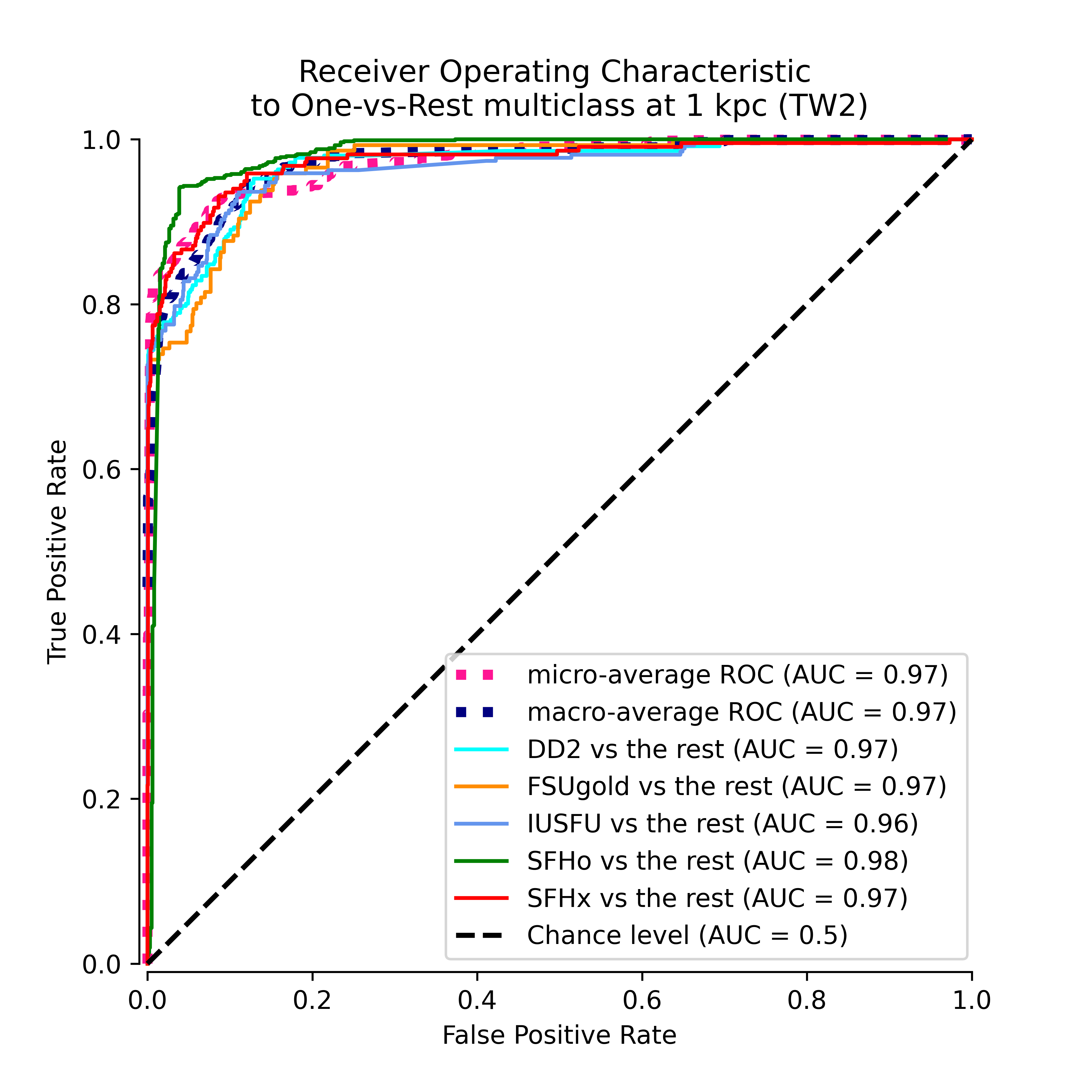}
		\includegraphics[width=0.32\textwidth]{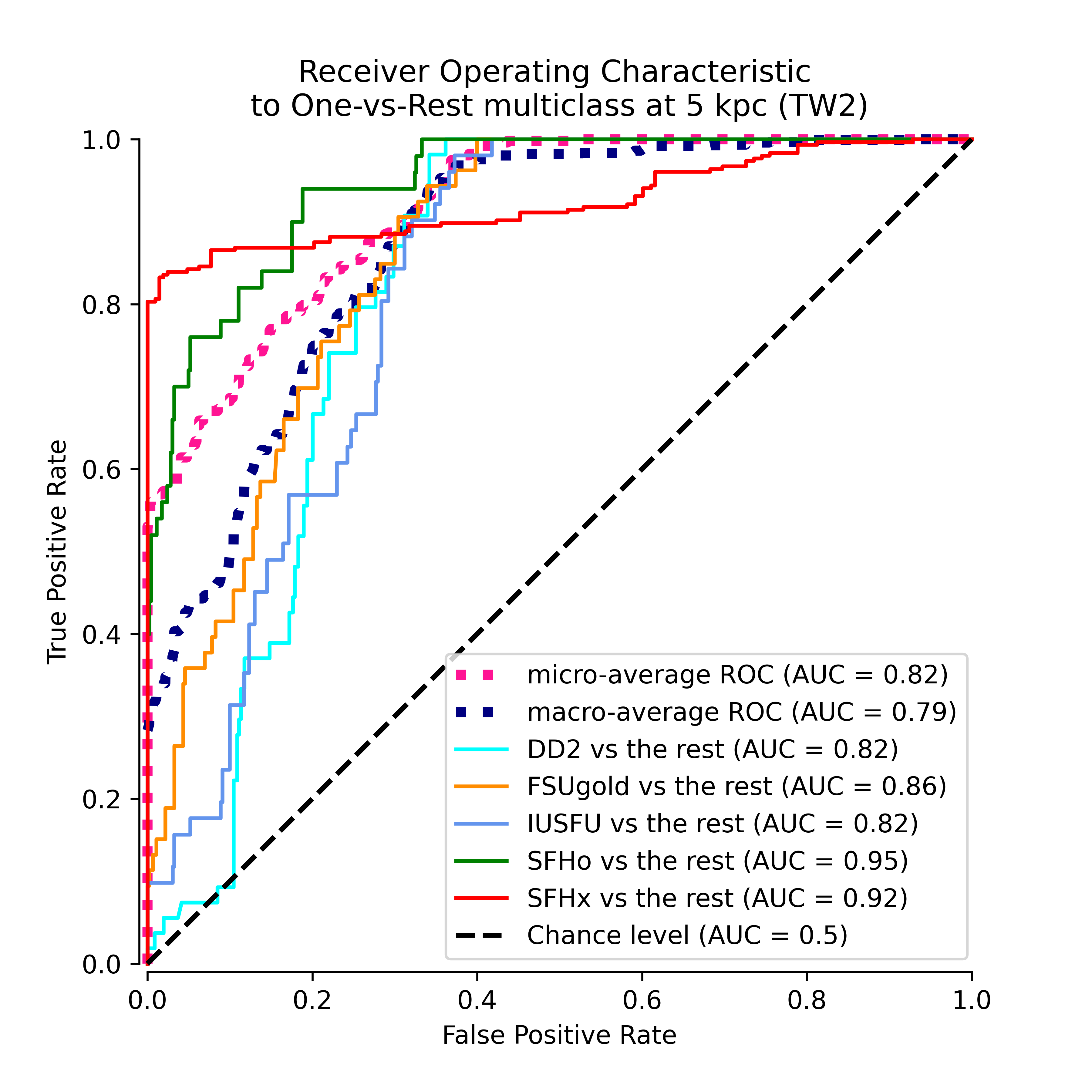}
		\includegraphics[width=0.32\textwidth]{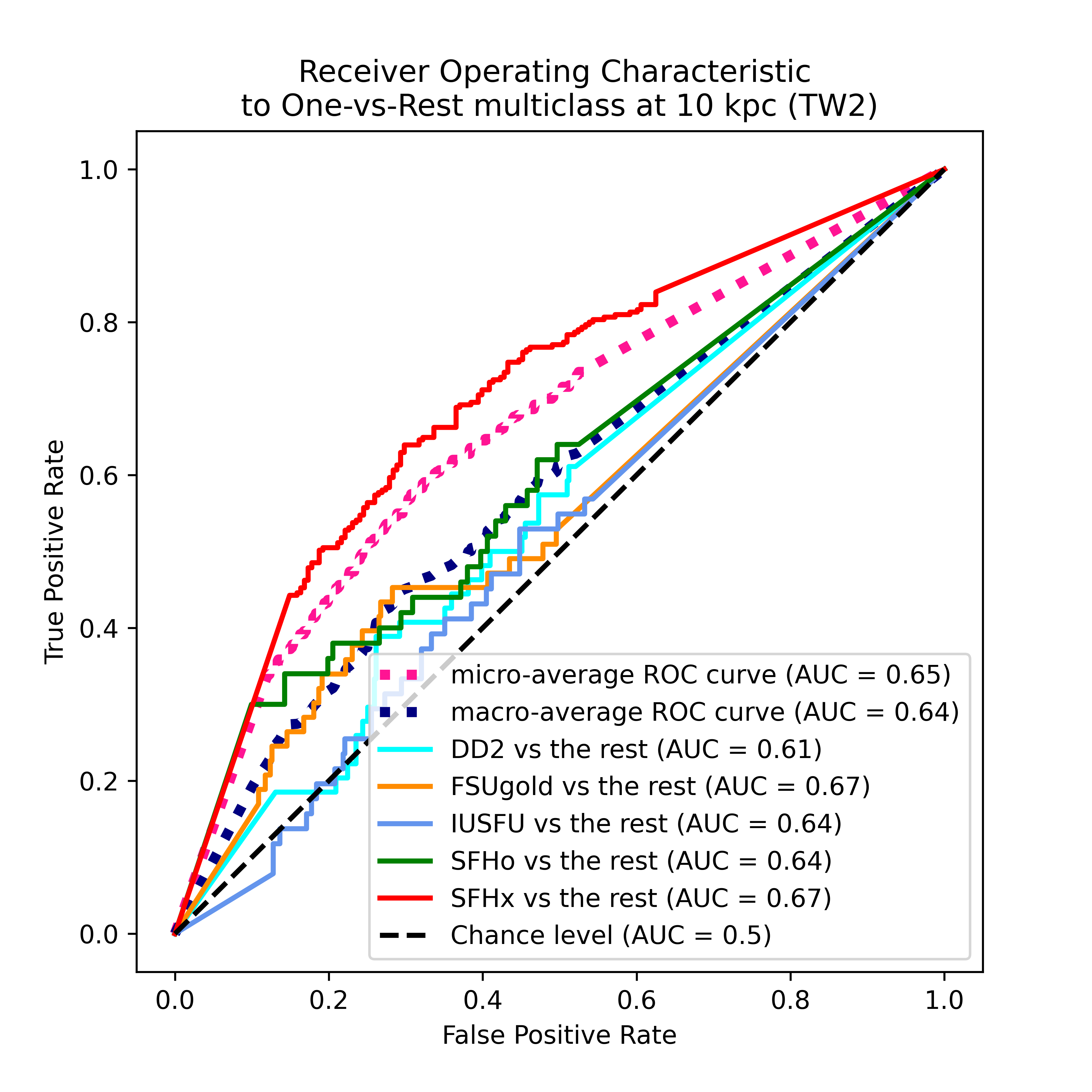}
		\includegraphics[width=0.32\textwidth]{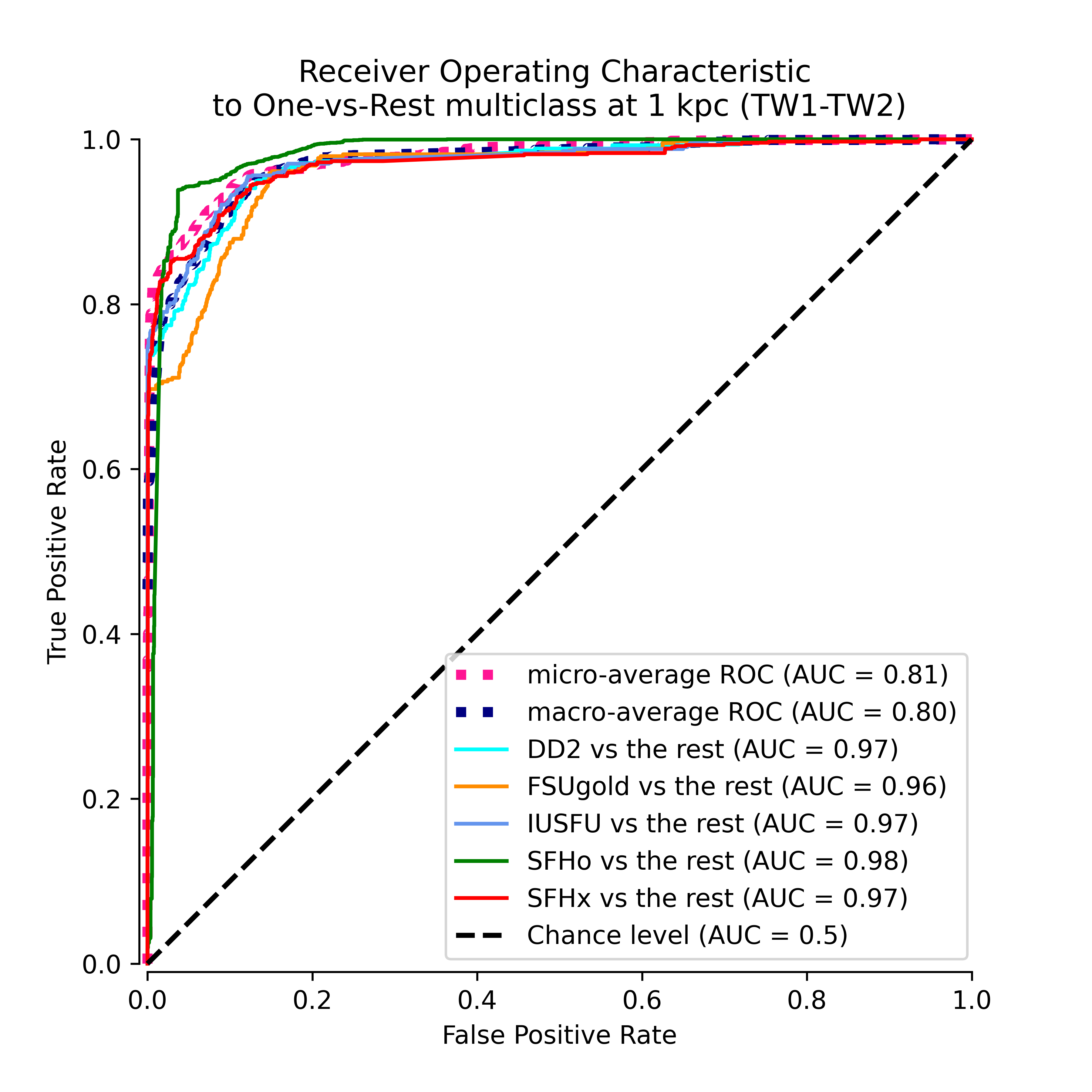}
		\includegraphics[width=0.32\textwidth]{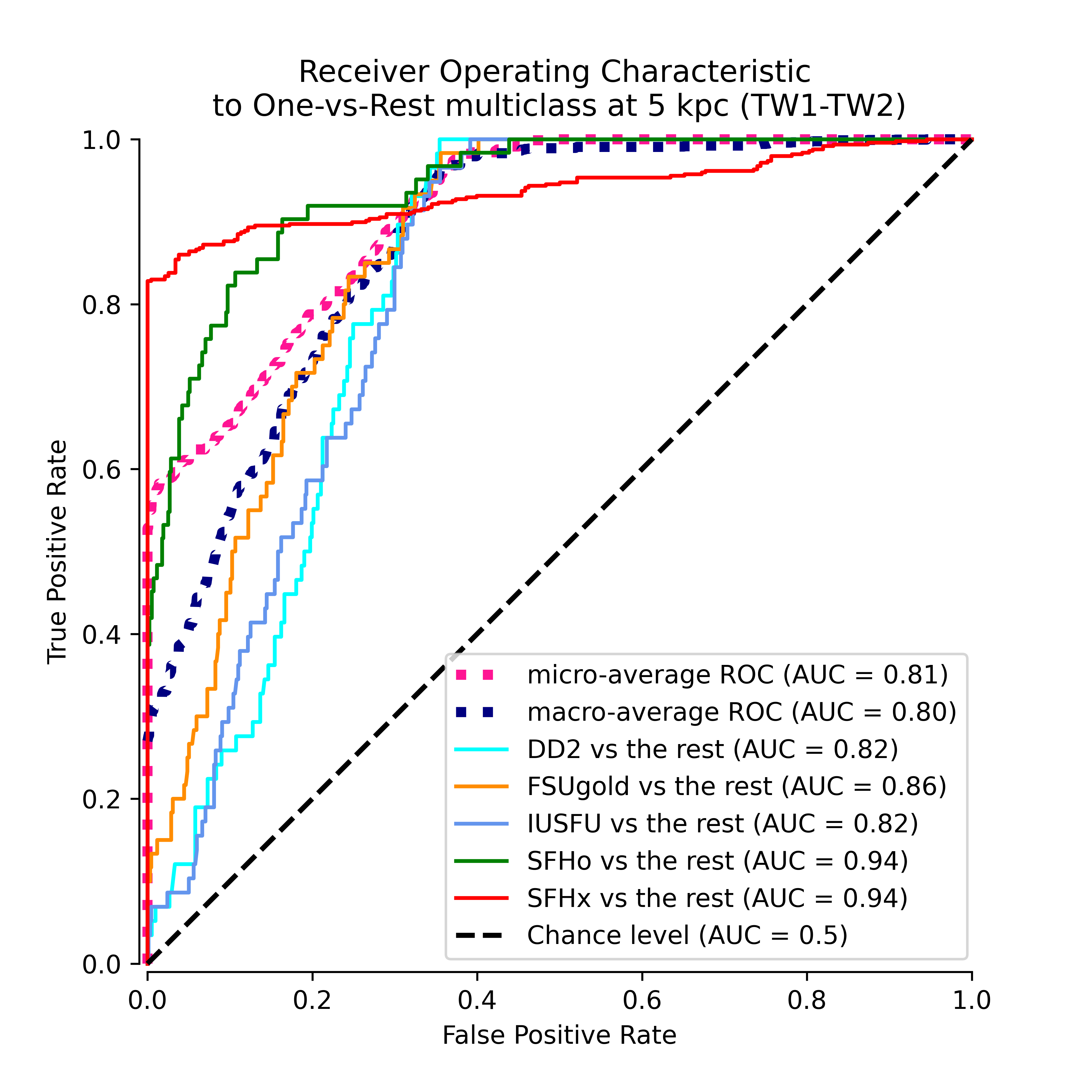}
		\includegraphics[width=0.32\textwidth]{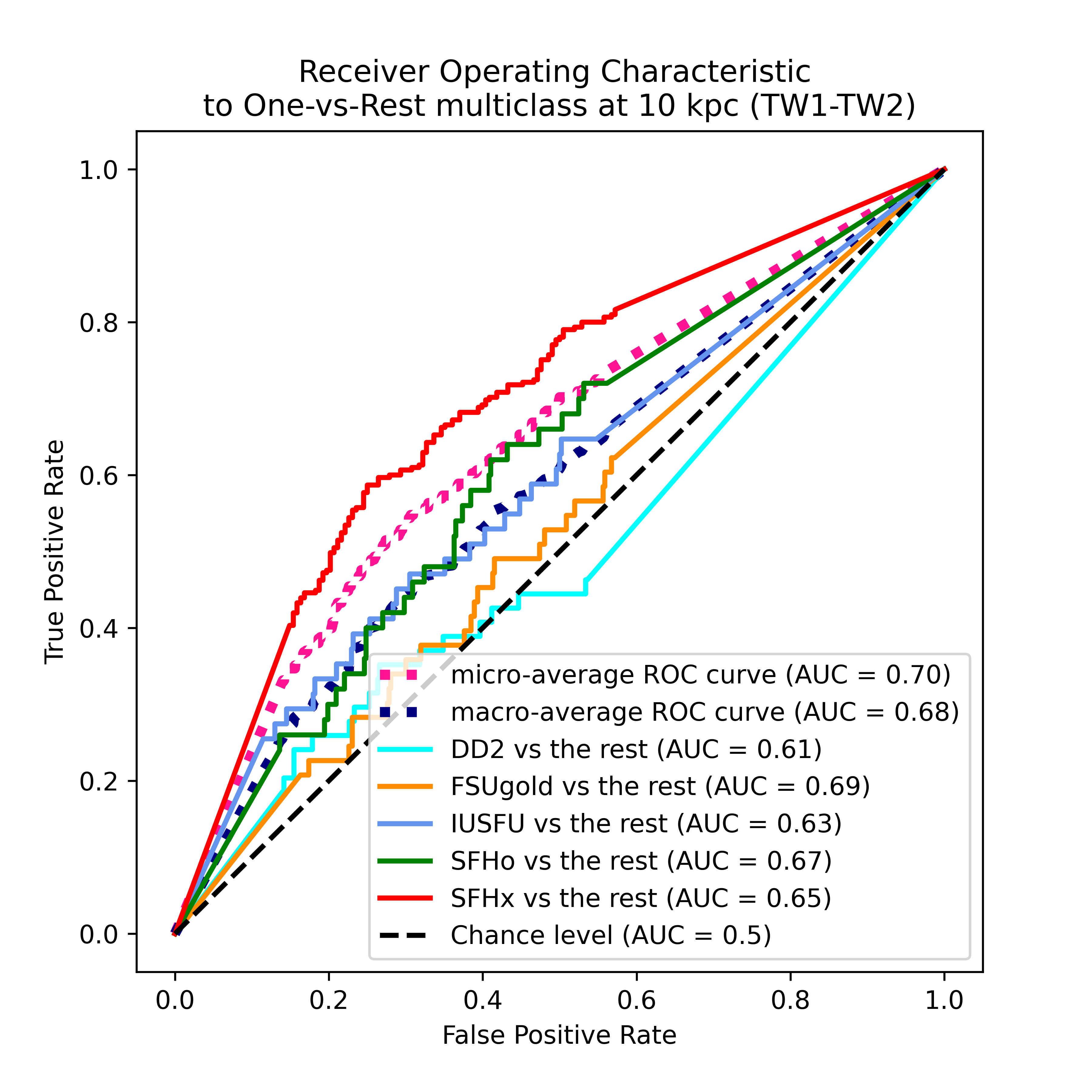}
		\caption{Receiver Operating Characteristic (ROC) curves and corresponding Area Under the Curve (AUC) values for EOS multiclassification across TW1, TW2, and TW1–TW2. Rows represent the time windows, while columns correspond to source distances of 1 kpc, 5 kpc, and 10 kpc. Results include individual EOS classes (one-vs-rest), as well as micro- and macro-average AUCs, obtained from the CNN model described in Section \ref{sec:Methodology}.}
		\label{fig:Results_ROC_TW1}
	\end{figure*}

	\textcolor{blue}{As illustrated by the ROC curves in Figure \ref{fig:Results_ROC_TW1}, the CNN model demonstrates a robust ability to distinguish between EOS classes at 1 kpc across all considered datasets.}
	\textcolor{blue}{First row in Figure \ref{fig:Results_ROC_TW1} shows that at 1 kpc, the CNN model demonstrates strong performance in distinguishing between EOS classes. This is evidenced by high ROC AUC scores for individual EOSs: DD2 (0.97), FSUgold (0.98), IUSFU (0.96), SFHo (0.98), and SFHx (0.97). Furthermore, the micro-average AUC of 0.89 and macro-average AUC of 0.80 provide a strong global indication of the model's efficacy in a one-vs-all multiclassification scheme. As expected, the model's ability to correctly classify the EOSs diminishes significantly at 10 kpc. 
		\\ \\
		The second and third rows of Figure \ref{fig:Results_ROC_TW1} demonstrate that the CNN model's performance in distinguishing between EOS classes at 1 kpc remains consistent across both the TW2 and TW1–TW2 datasets. This preservation of accuracy indicates that the model's efficacy is robust regardless of the specific time window used for analysis. Individual ROC (AUC) scores are high: DD2 (0.97), FSUgold (0.96), IUSFU (0.97), SFHo (0.98), and SFHx (0.97). The micro-average AUC of 0.81 and macro-average AUC of 0.80 further provide a strong global indication of the model's efficacy, comparable to the results obtained in studies 1 and 2.
		At 5 kpc, the model's recognition capabilities for different EOSs is reduced, as indicated by the ROC AUC scores: DD2 (0.82), FSUgold (0.86), IUSFU (0.82), SFHo (0.94), and SFHx (0.94). The micro-average AUC (0.81) and macro-average AUC (0.80) are very similar to the corresponding scores from previous studies, demonstrating sustained performance across different time periods for training and testing.
		At 10 kpc, the performance of the CNN model shows expected degradation with lower individual ROC AUC scores: DD2 (0.61), FSUgold (0.69), IUSFU (0.63), SFHo (0.67), and SFHx (0.65). The microaverage AUC (0.70) and macroaverage AUC (0.60) are indicative of the difficulties experienced in the operation of the algorithm to classify EOS at larger distances.}
	\section{Summary}
	\label{sec:summary}
	The design and implementation of the single-stack CNN architecture (described in Section \ref{sec:Methodology}) demonstrate the ability to correctly classify the five EOS classes, at 1 kpc,
	using the HFF estimated slope in LVK interferometric data. At 5 kpc and 10 kpc, the CNN model exhibits a degradation of its ability to differ between the distinct EOS classes.   
	Loss values show minor fluctuations or marginal increases both prior to and following the application of the SMOTE technique. In contrast, the micro-average and macro-average OvR (AUC) scores exhibit modest enhancements post-SMOTE, particularly at distances of 5 kpc and 10 kpc. This outcome suggests that, despite the inherent imbalanced characteristics observed in the datasets, the CNN algorithm maintains a consistent proportion of accurate classifications predicted by the model across the studies conducted, regardless of the number of instances present in the datasets.
	\\ \\
	The OvR (AUC) scores presented in Table \ref{tab:OvR_results} provide an evaluation of the CNN's ability to classify the five EOS classes at varying Galactic distances across the three studies developed. The distribution of macro-average AUC indicates that the application of the SMOTE technique has positively influenced the CNN's capability to classify the under-represented EOS classes. Specifically, the implementation of the SMOTE approach enables the inference of result consistency across all studies, thereby demonstrating the algorithm's robustness in classifying distinct EOS classes, regardless of the density of samples presented by the datasets and the temporal variations observed in TW1 and TW2.
	It also shows that the integration of combined data from both TW1 and TW2 provides an efficient training landscape, enhancing the CNN's ability to reliably classify all EOS types under challenging detection conditions involving time variations in LVK data.
	\begin{table}[!ht]
		\centering
		\caption{Accuracy, loss, micro-average OvR (AUC), and macro-average OvR (AUC) for the multi-classifier CNN before and after the application of the SMOTE technique on TW1  dataset.}
		\begin{adjustbox}{max width=\textwidth}
			\begin{tabular}{c| c@{\hspace*{1em}} c@{\hspace*{1em}} c@{\hspace*{1em}} c@{\hspace*{1em}} c@{\hspace*{1em}} c@{\hspace*{1em}}}
				\hline
				\hline
				\textbf{BEFORE SMOTE} &   &  &   \\
				\hline
				\hline
				Metric  & 1 kpc & 5 kpc & 10 kpc \\
				\hline
				\hline
				Accuracy                   & 0.93  & 0.82 &  0.73\\
				Loss                       & 0.32  & 0.43 &  0.57\\
				Micro-Average OvR (AUC)    & 0.89  & 0.89 &  0.73\\
				Macro-Average OvR (AUC)    & 0.80  & 0.80 &  0.68\\
				\hline
				\hline
				\textbf{AFTER SMOTE} & & &  \\
				\hline
				\hline
				Metric  & 1 kpc & 5 kpc & 10 kpc \\
				\hline
				\hline
				Accuracy                     & 0.95  & 0.83 &  0.74\\
				Loss                         & 0.36  & 0.47 &  0.59\\
				Micro-Average OvR (AUC)      & 0.91  & 0.89 &  0.65\\
				Macro-Average OvR (AUC)      & 0.82  & 0.81 &  0.64\\
				\hline
				\hline
			\end{tabular}
		\end{adjustbox}
		\label{tab:OvR_TW1}
	\end{table}
	\begin{table}[!ht]
		\centering
		\caption{Accuracy, loss, micro-average OvR (AUC), and macro-average OvR (AUC) for the multi-classifier CNN before and after the application of the SMOTE technique on TW2 dataset.}
		\begin{adjustbox}{max width=\textwidth}
			\begin{tabular}{c| c@{\hspace*{1em}} c@{\hspace*{1em}} c@{\hspace*{1em}} c@{\hspace*{1em}} c@{\hspace*{1em}} c@{\hspace*{1em}}}
				\hline
				\hline
				\textbf{BEFORE SMOTE} &   &  &   \\
				\hline
				\hline
				Metric  & 1 kpc & 5 kpc & 10 kpc \\
				\hline
				\hline
				Accuracy                   & 0.90  & 0.80 & 0.70\\
				Loss                       & 0.30  & 0.40 & 0.59\\
				Micro-Average OvR (AUC)    & 0.97  & 0.82 & 0.65\\
				Macro-Average OvR (AUC)    & 0.97  & 0.79 & 0.64\\
				\hline
				\hline
				\textbf{AFTER SMOTE} & & &  \\
				\hline
				\hline
				Metric  & 1 kpc & 5 kpc & 10 kpc \\
				\hline
				\hline
				Accuracy                     & 0.93  & 0.82  & 0.71\\
				Loss                         & 0.32  & 0.44  & 0.62\\
				Micro-Average OvR (AUC)      & 0.97  & 0.83  & 0.65\\
				Macro-Average OvR (AUC)      & 0.97  & 0.81  & 0.65\\
				\hline
				\hline
			\end{tabular}
		\end{adjustbox}
		\label{tab:OvR_TW2}
	\end{table}
	\begin{table*}
		\centering
		\caption{Receiver operating characteristic to One-vs-Rest area under the curve for multiclass classification, at Galactic distances of 1 kpc, 5 kpc, and 10 kpc, involving the three studies developed in this research.}
		\begin{adjustbox}{max width=\textwidth}
			\begin{tabular}{l@{\hspace*{1em}}|ccc|ccc|ccc} 
				\hline
				\hline
				\multicolumn{1}{c}{} & \multicolumn{1}{c}{} & \multicolumn{1}{c}{} & \multicolumn{1}{c}{} & \multicolumn{1}{c}{} & \multicolumn{1}{c}{\textbf{OvR (AUC) (\%)}} & \multicolumn{1}{c}{} & \multicolumn{1}{c}{} & \multicolumn{1}{c}{} & \multicolumn{1}{c}{}\\
				\hline
				\hline
				\multicolumn{1}{c}{} & \multicolumn{1}{c}{} & \multicolumn{1}{c}{\textbf{1 kpc}} & \multicolumn{1}{c}{} & \multicolumn{1}{c}{} & \multicolumn{1}{c}{\textbf{5 kpc}} & \multicolumn{1}{c}{} & \multicolumn{1}{c}{} & \multicolumn{1}{c}{\textbf{10 kpc}} & \multicolumn{1}{c}{} \\
				\multicolumn{1}{l}{CLASS} & \multicolumn{1}{c}{TW1} & \multicolumn{1}{c}{TW2} & \multicolumn{1}{c}{TW1-TW2} & \multicolumn{1}{c}{TW1} & \multicolumn{1}{c}{TW2} & \multicolumn{1}{c}{TW1-TW2} & \multicolumn{1}{c}{TW1} & \multicolumn{1}{c}{TW2} & \multicolumn{1}{c}{TW1-TW2}\\
				\hline
				\hline
				DD2      & 0.97 & 0.97 & 0.97 & 0.82 & 0.80 & 0.82 & 0.61 & 0.61 & 0.61 \\
				FSUgold  & 0.98 & 0.97 & 0.96 & 0.86 & 0.86 & 0.86 & 0.66 & 0.67 & 0.69 \\
				IUSFU    & 0.96 & 0.96 & 0.97 & 0.82 & 0.83 & 0.82 & 0.64 & 0.64 & 0.63 \\
				SFHo     & 0.98 & 0.98 & 0.98 & 0.95 & 0.94 & 0.94 & 0.68 & 0.64 & 0.67 \\
				SFHx     & 0.97 & 0.97 & 0.97 & 0.92 & 0.93 & 0.94 & 0.69 & 0.67 & 0.65 \\
				\hline
				\hline
			\end{tabular}
		\end{adjustbox}
		\label{tab:OvR_results}
	\end{table*}
	\textcolor{blue}{Our results, derived using noise characteristics consistent with the LVK O3b run, underscore the challenges and opportunities in GW data analysis. With the recent availability of data from the LIGO O4 observing run, characterized by improved detector sensitivity, we anticipate enhanced capabilities for estimating and classifying the HFF slope, particularly below 1 kHz. Furthermore, next-generation GW observatories, such as Cosmic Explorer \cite{Cosmic_Explorer} and the Einstein Telescope \cite{ET}, herald order-of-magnitude improvements in detection sensitivity. This study highlights the potential of using the CCSN GW HFF to enhance our comprehension of the intricate physical processes that control stellar collapse and the EOS.} 
	Given that GW signal amplitude scales inversely with distance, this increased sensitivity directly translates into an effectively greater detection range. Conservatively, the high classification performance achieved in our 1 kpc models could become representative of detections at 10 kpc or beyond with these future instruments. 
	\section*{Acknowledgements}
	This research has made use of data or software obtained from the Gravitational Wave Open Science Center (gwosc.org), a service of the LIGO Scientific Collaboration, the Virgo Collaboration, and KAGRA. This material is based upon work supported by NSF's LIGO Laboratory which is a major facility fully funded by the National Science Foundation, as well as the Science and Technology Facilities Council (STFC) of the United Kingdom, the Max-Planck-Society (MPS), and the State of Niedersachsen/Germany for support of the construction of Advanced LIGO and construction and operation of the GEO600 detector. Additional support for Advanced LIGO was provided by the Australian Research Council. Virgo is funded, through the European Gravitational Observatory (EGO), by the French Centre National de Recherche Scientifique (CNRS), the Italian Istituto Nazionale di Fisica Nucleare (INFN) and the Dutch Nikhef, with contributions by institutions from Belgium, Germany, Greece, Hungary, Ireland, Japan, Monaco, Poland, Portugal, Spain. KAGRA is supported by Ministry of Education, Culture, Sports, Science and Technology (MEXT), Japan Society for the Promotion of Science (JSPS) in Japan; National Research Foundation (NRF) and Ministry of Science and ICT (MSIT) in Korea; Academia Sinica (AS) and National Science and Technology Council (NSTC) in Taiwan. 
	A.C. and M.S. acknowledge Polish National Science Centre Grant No. UMO-2024/03/1/ST9/00005, and the Polish National Agency for Academic Exchange within Polish Returns Programme Grant No. BPN/PPO/2023/1/00019. 
	M.S. acknowledges also Polish National Science Centre Grant No. UMO-2023/49/B/ST9/02777. 
	M.Z. was supported in part by the National Science Foundation Gravitational Physics Experimental and Data Analysis Program through award PHY 2110555.
	AM and DM were supported in part by the National Science Foundation gravitational physics theory program through grant PHY-2409148.
	

	\appendix
	\section{Convolutional Neural Networks}\label{Ap:CNN_model}
	The architecture of the single CNN stack model implemented to classify the EOS by using the estimation of the HFF slope and the cWB event production is described in the following items
	\begin{itemize}
		\item [1.] \textbf{Input layer} \cite{ML1}: The input layer receives input images consisting of likelihood time-frequency maps $L$. Formally, these maps are represented as matrices of pixels $$\mathbf{L_i} \in \mathbb{R}^{N_\textrm{time} \times N_\textrm{freq} \times C},$$ where $N_\textrm{time}$ and $N_\textrm{freq}$ correspond to the number of rows and columns representing time and frequency dimensions, respectively, while $C$ denotes the channel. A channel designates a specific data dimension that encapsulates distinct types of information within single array of numbers. Typically, for input images, channels are associated with color components, with grayscale images containing a single channel. These maps are converted to high-resolution grayscale images and resized to dimensions $N_\textrm{time} = N_\textrm{freq}=28$. These images are fed directly into the initial layer for further processing. This data set, known as the training data set $D_\textrm{train}$, consists of all processed images derived from the cWB event production analysis. In subsequent stages, following the classification process, each image in the dataset will be categorized into one of five distinct classes: DD2, FSUgold, IUSFU, SFHo, and SFHx based on their HFF slopes, estimated from the second half of the LVK O3b scientific run using the methodology described in \cite{Casallas_2023} and \cite{Murphy_Casallas_2024}. 
		\item [2.] \textbf{Convolution Process} \cite{CNN1}: The convolutional layer applies filters or kernels $\mathbf{K} \in \mathbb{R}^{k_\textrm{time} \times k_\textrm{freq} \times C \times n}$ of dimensionality $k_\textrm{time}\times k_\textrm{freq}$, where $n$ represents the number of filters. These filters convolve over each image $L_i$, and the sliding kernels extract spatial features from the input data. The kernel is applied to each overlapping region of the input image, and for each position $(i,j)$, the pixel values are multiplied by the kernel weights, and subsequently these products are added to derive a single value. Each filter within a convolutional layer generates one feature map, and for $n$ filters, the layer produces $n$ feature maps that are stacked along the channel dimension. The convolution operation for the $n$-th filter, at position $(i,j)$ in the feature map, is represented by: 
		\begin{equation}\label{E:Conv}
			z_{i,j}^{(n)} = \sum_{c=0}^{C-1} \sum_{m=0}^{k_\textrm{time}-1} \sum_{n=0}^{k_\textrm{freq}-1} \mathbf{L}_{i \cdot s + m, j \cdot s + n, c} \cdot \mathbf{K}_{m,n,c}^{(n)} + b^{(n)},
		\end{equation}
		where $s$ denotes the stride (or the step size of the kernel $\mathbf{K}$), $b^{(n)}$ is the bias term for filter $n$, and the indices $i \cdot s + m$ and $j \cdot s + n$ account for the padding, $p$. Subsequently, a nonlinear function $f$ is applied to produce the final feature map:
		\begin{equation}\label{E:Feature}
			\mathbf{Z}_{i,j}^{(n)} = f\left(z_{i,j}^{(n)}\right).  
		\end{equation}
		In practical applications, feature maps are instrumental in detecting simple patterns, such as edges and corners. Feature maps in deeper layers integrate lower-level features into more intricate structures, such as shapes or objects. Furthermore, visualization of feature maps enables the interpretation of what the network ``perceives" at each layer. A pertinent observation is that, in contrast to DNN models, CNNs offer parameter efficiency by sharing weights across spatial positions, thereby mitigating overfitting.
		\item [3.] \textbf{ReLU Function} \cite{ML2}: The Rectified Linear Unit (ReLU) is a type of nonlinear activation function used extensively in neural networks. Within a CNN, the ReLU processes neurons from the convolutional layers, functioning to inject nonlinearity into the network. This allows for the learning of intricate patterns through operations that are performed element-wise, working independently on each value in input images. ReLU layers are generally located after convolutional layers and before grouping layers or additional convolutions. The ReLU function is mathematically expressed as:
		\begin{equation}\label{E:ReLU}
			f(\mathbf{Z}_{i,j,c}) = \max(0, \mathbf{Z}_{i,j,c})   
		\end{equation}
		where $\mathbf{Z}_{i,j,c}$ represents the preactivation value at the spatial position $(i,j)$ within the channel $C$. In the absence of nonlinear activations, sequential linear layers, such as convolutional layers, would degenerate to a mere linear transformation, hindering the network's capacity to learn complex functions. By nullifying negative inputs, ReLU leads to sparse data representations, potentially enhancing computational efficiency and diminishing overfitting. In contrast to saturating functions like \textit{sigmoid} or \textit{tanh}, ReLU maintains a non-decreasing gradient for positive inputs, facilitating the training of deep networks.
		
		\item [4.] \textbf{Max Pooling} \cite{CNN2}: 
		The pooling layer is a critical component in CNNs used to reduce the spatial dimensions (width and height) of input feature maps while retaining the most important information. This process, known as downsampling, involves sliding a two-dimensional filter over each channel of a feature map and summarizing the features within that region. The incorporation of pooling layers in the CNN architecture is essential for several reasons, it achieves dimensionality reduction, decreasing the number of parameters and computations, making the model faster and more efficient; it helps the network become invariant to small translations or distortions in the input image; it prevents overfitting by creating a more generalized representation; and it aids in building a hierarchical representation of features, where lower layers capture fine details and higher layers capture more abstract, global features. Ultimately, the stacking of convolution and pooling layers is a fundamental architectural pattern in typical CNN models. The max pooling is mathematically defined as follows:
		$$\mathbf{Z}_{i,j} = \max_{m,n \in \text{window}} \mathbf{X}_{i \cdot s + m, j \cdot s + n}.$$
		$Z_{i,j}$ represents a single value in the output feature map at position $(i,j)$ after the pooling operation. $\max_{m,n \in \text{window}}$ is the operation that selects the maximum value from the elements within a specific window of the input. The variables $m$ and $n$ are the indices that iterate through the rows and columns of this window. $X$ is the input feature map. The indices $i\cdot s+m$ and $j\cdot s+n$ specify the exact location of the element being considered within this input map. Finally, $s$, is the stride, which dictates how many steps the pooling window shifts across the input feature map after each computation.
		\item [5.] \textbf{Fully-Connected}: A fully connected layer (FC) within a CNN acts as the final phase to merge features and make decisions. It integrates the high-level features identified by earlier convolutional or pooling layers into either class scores or regression results. The FC layer receives a 3D tensor as input from the final convolutional/pooling layer, with dimensions $H \times W \times C$. By flattening, this 3D tensor is transformed into a 1D vector:  
		$\mathbf{x} \in \mathbb{R}^{D}, \quad D = H \times W \times C.$
		For an FC layer containing $N$ neurons, the transformation is 
		$\mathbf{y} = \mathbf{W} \mathbf{x} + \mathbf{b},$
		where  
		$\mathbf{W} \in \mathbb{R}^{N \times D}$ represents the weight matrix,   
		$\mathbf{b} \in \mathbb{R}^{N}$ is the bias vector, and $\mathbf{y} \in \mathbb{R}^{N}$ is the output logits (pre-activation stage). 
		\item [6.] \textbf{Softmax} \cite{ML2}: The Softmax Layer is an essential component in CNNs used for classification tasks. It transforms raw output scores, known as logits, from the network into probabilities, thereby allowing the model to allocate confidence scores to each class. A vector of logits is obtained from the final fully connected (FC) layer, with $$\mathbf{z} = [z_1, z_2, ..., z_K] \in \mathbb{R}^K,$$ representing the number of classes. For each class $i$, the probability $p_i$ is calculated as $$p_i = \frac{e^{z_i}}{\sum_{j=1}^K e^{z_j}},$$ ensuring both normalization $\sum_{i=1}^K p_i = 1$ and probabilistic interpretation $0 \leq p_i \leq 1$. The purpose of the \textit{softmax} function is to enhance interpretability by converting abstract logits into probabilities. It facilitates the generation of class confidence, wherein the class with the highest probability is deemed the model's prediction, and provides differentiability that supports gradient-based optimization techniques, such as backpropagation.
		\item [7.] \textbf{Loss Function}\cite{ML1}: A loss function is a mathematical tool used to measure the difference between a model's predictions and the actual target values evaluating how closely these predictions align with the real output. In the context of CNNs the loss function similarly assesses the error between predicted outcomes and the true labels. This function acts as a training signal for fine-tuning the model's parameters through gradient-based techniques such as backpropagation.
		A loss function is a scalar-valued function that translates the relationship between the model's predictions and the actual labels into a single real number. Formally, it is expressed as: $$\mathcal{L}: \mathcal{Y} \times \mathcal{Y} \rightarrow \mathbb{R},$$ where
		$\mathcal{Y}$ represents the set of predictions (e.g., class probabilities, regression outcomes) and $\mathcal{Y}$ denotes the set of true labels.  
		Given a data set $\{(\mathbf{x}_i, \mathbf{y}_i)\}_{i=1}^N$, the overall loss is usually calculated as the average across all samples:
		$\mathcal{L}(\theta) = \frac{1}{N} \sum_{i=1}^N \ell(f_\theta(\mathbf{x}_i), \mathbf{y}_i) + \lambda \cdot \text{Reg}(\theta),$
		where $\ell$ is the per sample loss (e.g., cross-entropy, MSE).  
		$f_\theta$ represents the model with parameters \( \theta \).  
		$\text{Reg}(\theta)$ indicates the regularization component (e.g., L1/L2 penalties), and $\lambda$ is the regularization strength.
		\item [8.] \textbf{Learning} \cite{CNN_GW_3}: We focus on multi-class classification, meaning there are multiple classes, but each object belongs to a single class. Our objective is to use the training dataset to determine the characteristics of each class. This process is known as classifier training or model learning. The output generated by the model is a label or class assigned to each sample within the training dataset. Specifically, our classifier will be capable of distinguishing among five different EOS based on the HFF slope values once they are estimated in different realizations of LVK noise.
		\item [9.] \textbf{Evaluation and Validation} \cite{Antelis_2022, CNN1}: We evaluated the efficacy of our classifier by employing a Monte Carlo method, in which the dataset is divided with 70\% allocated for training and the remaining 30\% allocated for testing. This approach is implemented using two distinct time windows, TW1 and TW2 [see table \ref{tab:Time_Window}], which are associated with the LVK Scientific Run O3b. To further evaluate the classifier's performance, we employ a Monte Carlo method that involves training with time window TW1 while testing with time window TW2. Throughout these validation processes, we derive the confusion matrix, as well as the micro and macro average (OvR) ROC, in addition to the ROC for each class within the two time windows. 
	\end{itemize}
	\section{Performance metrics}\label{Ap:Performance}
	The CNN model's effectiveness is validated on previously unseen data; therefore, accurately evaluating performance metrics during testing is crucial. We used the following metrics to evaluate the ability of the proposed model to classify the EOS.
	
	\textbf{(i) Confusion Matrix}\cite{CNN1}: The confusion matrix stands as a fundamental tool for systematically evaluating the performance of classification models. This tabular representation offers a visualization that directly compares the model's predicted class labels against the ground truth labels present in the dataset \cite{ML1}, providing a detailed breakdown of both correct and incorrect classifications, 
	For binary classification problems, the confusion matrix is a $2\times 2$ table comprising four key metrics: True Positives (TP), True Negatives (TN), False Positives (FP), and False Negatives (FN). In the context of multiclass classification problems, the confusion matrix naturally extends from this $2\times 2$ structure to an $n\times n$ structure, where $n$ represents the total number of distinct classes \cite{CNN2}. In this expanded matrix, each row conventionally represents the actual (true) class of the instances, while each column corresponds to the class predicted by the model. The diagonal elements of the confusion matrix indicate the number of instances that were correctly classified for each specific class. Conversely, the off-diagonal elements represent misclassifications, meaning, the number of instances that truly belonged to a specific class but were incorrectly predicted as belonging to a different one. Analyzing these off-diagonal values is relevant for identifying specific types of errors the model makes (e.g., discerning which classes are frequently confused with others), thereby providing actionable insights for model refinement and improvement.
	
	\textbf{(ii) The AUC-ROC Curve}\cite{Antelis_2022}: The Area Under the Curve (AUC) Receiver Operating Characteristics (ROC) curve is a widely accepted graphical tool for evaluating the performance of classification models. It illustrates the ability of a classifier system as its discrimination threshold is varied. Specifically, the ROC curve plots the TPR, also known as sensitivity or recall, against the False Positive Rate (FPR) for different threshold settings.
	\\ \\
	Although fundamentally designed for binary classification, ROC analysis can be extended to evaluate models in multiclass classification problems through established strategies. The most common and intuitive approach is the One-vs-Rest (OvR) (also known as One-vs-All, OvA) method. This technique effectively decomposes the multiclass prediction task into a series of independent binary classification problems. For each class in the multiclass set, the OvR strategy treats that specific class as the ``positive" class and groups all other classes as the ``negative" class.
	For an N-class problem, the OvR approach thus generates N distinct binary classifiers. An independent ROC curve and a corresponding AUC score are computed for each of these N binary classification problems. Each individual ROC curve and its AUC value then specifically illustrate the model's ability to discriminate its respective "positive" class from all other classes combined. Higher AUC scores for individual classes indicate a more robust ability of the model to correctly identify instances of that specific class while correctly rejecting instances belonging to any of the other classes. This detailed per-class analysis is crucial for understanding specific strengths and weaknesses of the multiclass model across its different output categories.
	
	\textbf{(iii) Macro and Micro Average}\cite{ML2}: In the evaluation of multi-class classification models, particularly in ML and pattern recognition, the choice between micro- and macro-averaging for performance metrics is crucial and depends heavily on the characteristics of the dataset. These two averaging techniques provide distinct perspectives on model performance, especially when dealing with imbalanced datasets.
	The macro-averaging can be considered as a ``class-centric evaluation" meaning it computes a given performance metric independently for each class and then calculates the arithmetic mean of these per-class scores. Formally, if we have $C$ classes and $M_i$ is the metric score for class $i$, the macro-averaged metric $M_\textrm{macro}$ is given by
	\begin{equation}
		M_{\textrm{macro}} = \frac{1}{C} \sum_{i=1}^{C} M_i\;.
	\end{equation}
	The fundamental principle of macro-averaging is to assign equal weight to each class, regardless of its prevalence in the dataset. This means that the performance on a minority class contributes as much to the overall macro-average score as the performance on a majority class.
	Consequently, macro-averaging is highly sensitive to the model's performance on minority classes. A poor performance on even a single rare class can significantly depress the macro-averaged score. This makes it an ideal choice when all classes are considered equally important from an application standpoint, and misclassification of minority instances is as critical as misclassification of majority instances.
	Macro-averaging is particularly valuable in the presence of class imbalance, where the number of instances varies significantly across different class labels. In such scenarios, if a model performs poorly on under-represented classes, the macro-average will reflect this deficiency, providing a more realistic assessment of the model's generalized performance across all categories, rather than being skewed by high accuracy on over-represented classes.
	\\ \\
	Micro-averaging, conversely, aggregates the counts of TP, FP, and false negatives (FN) across all classes first, and then calculates the performance metric based on these global sums. The micro-averaged precision ($P_{\textrm{micro}}$) and recall ($R_{\textrm{micro}}$) are defined as follows:
	\begin{equation}
		P_{\textrm{micro}} = \frac{\sum_{i=1}^{C} TP_i}{\sum_{i=1}^{C} TP_i + \sum_{i=1}^{C} FP_i}
	\end{equation}
	\begin{equation}
		R_{\textrm{micro}} = \frac{\sum_{i=1}^{C} TP_i}{\sum_{i=1}^{C} TP_i + \sum_{i=1}^{C} FN_i}
	\end{equation}
	Micro-averaging effectively treats every individual instance equally, regardless of its class label. This means that the performance on larger classes will have a disproportionately greater influence on the micro-averaged score compared to smaller classes.
	In the presence of class imbalance, micro-averaging will be dominated by the performance on the majority classes. A model that performs exceptionally well on large classes but poorly on small classes might still achieve a high micro-averaged score. This can be misleading if the objective is to ensure robust performance across all categories.
	Micro-averaging is appropriate when the primary concern is the overall correctness of predictions across the entire dataset, irrespective of class distribution. It is often used when the goal is to optimize for global performance and when errors in majority classes are considered more impactful due to their sheer number (e.g., in very large-scale classification systems where overall throughput accuracy is paramount, or when the cost of misclassification is uniform across instances).

	

\end{document}